\documentclass [fontsize=11, headings=normal, headsepline=true,parskip=half]{article}
\usepackage{amsmath,mathtools}
\usepackage{scrpage2}
\usepackage {blindtext}
\usepackage[latin1]{inputenc}
\usepackage[T1]{fontenc}
\usepackage{color} 
\usepackage{placeins}
\usepackage{url}
\usepackage[arrow, matrix, curve]{xy}
\usepackage[symbol]{footmisc}
\usepackage{graphicx}
\usepackage{dsfont}
\usepackage{wrapfig}
\usepackage{bbm}
\usepackage{paralist} 
\usepackage{amssymb}
\usepackage{stmaryrd}
\usepackage{float}
\usepackage[dvipsnames]{xcolor}
\usepackage{pdfsync}
\usepackage{pdfpages}
\usepackage{booktabs}
\usepackage{nicefrac}      
\usepackage{microtype}    
\usepackage{algorithm}
\usepackage[noend]{algpseudocode}
\algrenewcommand\alglinenumber[1]{\scriptsize #1:}
\usepackage[numbers]{natbib}
\usepackage{hyperref}
\hypersetup{
	colorlinks,
	citecolor=black,
	filecolor=black,
	linkcolor=blue,
	urlcolor=black
}

\newcommand{\Var}{\operatorname{Var}}

\newcommand{\EW}{\mathbb{E}}

\newcommand{\iid}{\stackrel{iid}{\sim}}

\newcommand{\R}{\mathbb{R}}

\newcommand{\bfX}{{\mathbf X}}

\newtheorem{lemma}{Lemma}

\title{Adaptive penalization in high-dimensional regression and classification with external covariates using variational Bayes}	
\author{Britta Velten\\
		European Molecular Biology Laboratory\\
		Heidelberg, Germany\\
		\texttt{britta.velten@embl.de}\\
		\and\\
		Wolfgang Huber\\
		European Molecular Biology Laboratory\\
		Heidelberg, Germany\\
		\texttt{wolfgang.huber@embl.de}\\
		}

\date{}
\begin{document}
	\maketitle

\begin{abstract}
Penalization schemes like Lasso or ridge regression are routinely used to regress a response of interest on a high-dimensional set of potential predictors. Despite being decisive, the question of the relative strength of penalization is often glossed over and only implicitly determined by the scale of individual predictors. At the same time, additional information on the predictors is available in many applications but left unused. Here, we propose to make use of such external covariates to adapt the penalization in a data-driven manner. We present a method that differentially penalizes feature groups defined by the covariates and adapts the relative strength of penalization to the information content of each group.  Using techniques from the Bayesian tool-set our procedure combines shrinkage with feature selection and provides a scalable optimization scheme. 

We demonstrate in simulations that the method accurately recovers the true effect sizes and sparsity patterns per feature group. Furthermore, it leads to an improved prediction performance in situations where the groups have strong differences in dynamic range. In applications to data from high-throughput biology, the method enables re-weighting the importance of feature groups from different assays. Overall, using available covariates extends the range of applications of penalized regression, improves model interpretability and can improve prediction performance.

\textbf{Code Availability} The software is freely available as an R package \url{https://git.embl.de/bvelten/graper}, scripts for the analyses contained in this paper can be found at \url{https://git.embl.de/bvelten/graper_analyses}.
\end{abstract}

\newpage
\section{Introduction} 
We are interested in the setup where we observe a continuous or categorical response $Y$ together with a vector of potential predictors, or features, $X \in \R^p$  and aim to find a relationship of the form	
$$ Y = f(X).$$
Two main questions are of potential interest in this setting. 
First, we want to obtain an $f$ that yields good predictions for $Y$ given a new observation $X$. Second, we aim at finding which components in $X$ are the 'important ones' for the prediction.
 
A common and useful approach to this end are (generalized) linear regression methods, which assume that the distribution of $Y|X$ depends on X via a linear term $X^T\beta$. In order to cope with high-dimensionality of $X$ and avoid over-fitting, penalization on $\beta$ is employed, e.g. in ridge regression \citep{hoerl1970ridge}, Lasso \citep{tibshirani1996regression} or elastic net \citep{zou2005regularization}. By constraining the values of $\beta$, the complexity of the model is restricted, resulting in biased but less variable estimates and improved prediction performance. In addition, some choices of the penalty yield estimates with a relatively small number of non-zero components, thereby facilitating feature selection. An example is the $L_1$-penalty employed in Lasso or elastic net.  

Commonly, penalization methods apply a penalty that is symmetric in the model coefficients. Real data, however, often consists of a collection of heterogeneous features, which such an approach does not account for. In particular, it ignores any additional information or structural differences that may be present in the features. Often we encounter $X$ whose components comprise multiple data modalities and data qualities, e.g., measurement values from different assays. Other side-information on individual features could include temporal or spatial information, quality metrics associated to each measurement or the features' sample variance, frequency or signal-to-noise ratio.  It has already been observed in multiple testing that the power of the analysis can be improved by making use of such external information (e.g. \citep{ignatiadis2016data, ferkingstad2008unsupervised, dobriban2015optimal, li2016multiple, lei2018adapt}). However, in current penalized regression models this information is frequently ignored. Making use of it could on one hand improve prediction performance. On the other hand, it might yield important insight into the relationship of external covariates to the features' importance. For example, if the covariate encodes different data modalities, insights into their relative importance could help cutting costs by reducing future assays to the essential data modalities.

As a motivating example we consider applications in molecular biology and precision medicine. Here, the aim is to predict phenotypic outcomes, such as treatment response, and identify reliable disease markers based on molecular data.  Nowadays, different high-throughput technologies can  be combined to jointly measure thousands of molecular features from different biological layers \citep{ritchie2015methods, hasin2017multi}. Examples include genetic alterations, gene expression, methylation patterns, protein abundances or microbiome occurrences. However, despite the increasing availability of molecular and clinical data, outcome prediction remains challenging \citep{alyass2015big,chen2013promise, hamburg2010path}. Common applications of penalized regression only make use of parts of the available data. For example, different assay types are simply concatenated or analysed separately.  In addition, available annotations on individual features are left unused, such as its chromosomal location or gene set and pathway membership. Incorporating side-information on the assay type and spatial or functional annotations could help to improve prediction performance. Furthermore, it could help prioritizing feature groups, such as different assays or gene sets.

Here, we propose a method that incorporates external covariates in order to guide penalization and can learn relationships of the covariate to the feature's effect size in a data-driven way. We introduce the method for linear models and extend it to classification purposes. We demonstrate that this can improve prediction performance and yields insights into the relative importance of different feature sets, both on simulated data and applications in high-throughput biology.

\section{Methods}
\label{methodsSec}
\subsection{Problem statement}
\label{problem_outline}
 Assume we are given observations $(x_1, y_1), \dots, (x_n, y_n)$ with $y_i \in \mathcal{Y} \subseteq \mathbb{R}$, $x_i \in  \mathbb{R}^p$ (possibly $n\ll p$) from a linear model, i.e.
	\begin{equation}
	y_i =  x_i^T\beta +\epsilon_i
	\end{equation}
with $\epsilon_i \iid \text{N}(0, \sigma^2)$. In addition, we suppose that we have access to a covariate $\zeta_j \in \mathcal{Z} \subseteq \R^k$ for each predictor $j=1,\dots, p$. We hope, loosely speaking, that $\zeta_j$ contains some sort of information on the magnitude of $\beta_j$. The question we want to address is: Can we use the information from $\zeta$ to improve upon estimation of $\beta$ and prediction of $Y$?

In order to estimate $\beta$ from a finite sample $y=(y_i)_{i=1,\dots,n} \in \R^{n}$ and $\mathbf{X} = [x_1,\dots, x_n]^T \in \R^{n \times p}$  we can employ penalization on the negative log-likelihood of the model, i.e.
\begin{equation}
\hat{\beta}(\lambda) \in \arg \min_\beta \frac{1}{n}||y - \mathbf{X}\beta||_2^2 + \lambda p(\beta),
\label{eq:pen}
\end{equation}
where $p$ denotes a penalty function on the model coefficients. For example, $p(\beta) = \sum_j |\beta_j|^q$ leads to Lasso ($q=1$) or ridge regression ($q=2$). The parameter $\lambda$ controls the amount of penalization and thereby the model complexity. Ideally, we would like to choose an optimal $\lambda$. For estimation this means minimizing the mean squared error $\text{MSE}(\hat{\beta}(\lambda)) = \EW ||\hat{\beta}(\lambda) - \beta ||_2^2$; for prediction this means minimizing the expected prediction error. In practice, $\lambda$ is often chosen to minimize the cross-validated error.

In most applications, the penalization is symmetric, i.e. for any permutation $\pi$ we have $\lambda p(\beta_1,\dots, \beta_p) = \lambda p(\beta_{\pi(1)}, \dots, \beta_{\pi(p)})$. However, as we have external information on each feature given by $\zeta$ we want to allow for differential penalization guided by $\zeta$. For this, we will consider the following non-symmetric generalization, which still leads to a convex optimization problem in $\beta$ for convex penalty functions $\tilde{p}$, such as $\tilde{p}(x) = |x|$ or $\tilde{p}(x) = x^2$:
		\begin{equation}
		\hat{\beta}(\lambda) \in \arg \min_\beta \frac{1}{n}||{y} - \mathbf{X}\beta||_2^2 + \sum_j \lambda(\zeta_j) \tilde{p}(\beta_j).
		\label{eq:diffPen}
		\end{equation} 
	 Instead of a constant $\lambda$, here $\lambda: \mathcal{Z} \rightarrow \R_{\geq0}$ provides a mapping from the covariate $\zeta$ to a non-negative penalty factor $\lambda(\zeta)$. This additional flexibility compared to a single penalty parameter can be helpful if $\zeta$ contains information on $\beta$. For example, in the simple case of ridge regression with deterministic orthonormal design matrix, known noise variance $\sigma^2$ and 'oracle covariate' $\zeta_j = \beta_j$ the optimal $\lambda$ is  seen to be $\lambda^*(\zeta_j) = \frac{\sigma^2}{\zeta_j^2}$. However, in practice the information in $\zeta$ is not that explicit and hence we do not know which $\lambda$ is optimal. 
	 
	 If $\lambda$ takes values in a small set of discrete values, e.g. for categorical covariates $\zeta$, cross-validation could be used to determine a suitable set of function values. This approach is employed in \citep{boulesteix2017ipf}, where categorical covariates encode different data modalities. However, cross-validation soon becomes prohibitive, as it requires a grid search exponential in the number of categories defined by $\zeta$. Similarly, cross-validation can be employed with $\lambda$ parametrized by a small number of tuning parameters using domain knowledge to come up with a suitable parametric form for $\lambda$ \citep{bergersen2011weighted,verissimo2016degreecox}. However, such an explicit form is often not available. In many situations it is a major problem itself to come up with a helpful relationship between $\zeta$ and $\beta$ and thereby knowledge of which values of a covariate would require more or less penalization. Therefore, we aim at finding $\lambda$ in a data-driven manner and with improved scalability compared to cross-validation.

		\subsection{Problem statement from a Bayesian perspective}
	\label{BayesPersp}
		
		There is a direct correspondence between estimates obtained from penalized regression and a Bayesian estimate with penalization via corresponding priors on the coefficients. For example, the ridge estimate corresponds to the maximum a posterior estimate (MAP) in a Bayesian regression model with normal prior on $\beta$ and the Lasso estimate to a MAP with a Laplace prior on $\beta$.  This correspondence opens up alternative strategies using tools from the Bayesian mindset to approach the problem outlined above: Differential penalization translates to introducing different priors on the components of $\beta$. Our belief that $\zeta$ carries information on $\beta$ can be incorporated by using prior distributions whose parameters depend on $\zeta$. In \citep{wiel2016better}, the authors used this idea to derive an Empirical Bayes approach for finding group-wise penalty parameters in ridge regression. However, this approach does not obviously generalize to other penalties such as the Lasso. 
		
		Moving completely into the Bayesian mindset we instead turn to explicit specification of priors to implement the penalization task. Different priors have been suggested \citep{mitchell1988bayesian, park2008bayesian, mackay1996bayesian, carvalho2009handling} and structural knowledge was incorporated into the penalization by employing multivariate priors that encode the structure in the covariance or non-exchangeable priors with different hyper-parameters (e.g. \citep{xu2015bayesian, wu2014sparse, andersen2015bayesian, hernandez2013generalized,rockova2014incorporating, hernandez2013generalized, engelhardt2014bayesian} and references therein). Despite the possible gains in prediction performance when incorporating such structural knowledge, these methods have not been widely applied. A limiting factors has often been the lack of scalability to large data sets.

\label{ProbOutline}
\subsection{Setup and notation}
\label{modeldef}
From the linear model assumption we have
\begin{equation}
y_i=x_i^T\beta +\epsilon_i \qquad \epsilon_i \iid \text{N}(0,{\tau}^{-1}),
\end{equation}
where $\tau$ denotes the precision of the noise.
 Based on the external covariate $\zeta$  we define a partition of the $p$ predictors into $G$ groups:
\begin{equation}
g_\zeta = g:\{1,\dots, p\} \rightarrow \{1,\dots, G\}.
\end{equation}
For instance, categorical covariates $\zeta$, such as different assay types, naturally define such a partition. For continuous covariates $g_\zeta$ can be defined based on suitable binning or clustering.

To achieve penalization in dependence of $\zeta$ we consider a spike-and-slab prior \citep{mitchell1988bayesian} on the model coefficients $\beta$ with a different slab precision $\gamma$ and mixing parameter $\pi$ for each group. We re-parametrize $\beta$ as $\beta_j=s_j b_j$ with
\begin{align}
b_j|\gamma_{g_\zeta(j)} &\sim \text{N}\left(0,\gamma_{g_\zeta(j)}^{-1}\right), \label{ss-normal}
\\
s_j|\pi_{g_\zeta(j)}&\sim \text{Ber}(\pi_{g_\zeta(j)}).
\label{ss-ber}
\end{align}
In the special case of $\pi=1$ this yields a normal prior as in \citep{mackay1996bayesian} corresponding to ridge regression. With $\pi<1$ we additionally promote sparsity on the coefficients, and the value of $\pi$ controls the number of active predictors in each group. The value of $\gamma$ controls the overall shrinkage per group.
To learn the model hyperparameters $\gamma$, $\pi$ and the noise precision $\tau$, we choose the following conjugate priors
\begin{align}
\tau &\sim\Gamma(r_\tau, d_\tau),\\
\intertext{and for each group $k \in \{1,\dots, G\}$}
\gamma_k &\sim \Gamma(r_\gamma, d_\gamma),\\
\pi_k &\sim \operatorname{Beta}(d_\pi, r_\pi),
\end{align}
with $d_\tau,r_\tau, d_\gamma , r_\gamma = 0.001 $ and $r_\pi, d_\pi = 1$.
Hence, the joint probability of the model is given by
\begin{align}
p(y,b,s,\gamma,\pi,\tau)=p(y|b,s, \tau) p(b,s|\pi,\gamma)p(\gamma)p(\pi)p(\tau).
\label{eq:model}
\end{align}

\subsection{Inference using Variational Bayes}
\label{inference}
The challenge now lies in inferring the posterior of the model parameters from the observed data  $\bfX,y$ and the covariate $\zeta$. While Markov Chain Monte Carlo methods are frequently used for this purpose they do not scale well to large data sets. Here, we adopt a variational inference framework \citep{blei2017variational,bishop2006pattern} that has been used (in combination with importance sampling) for variable selection with exchangeable priors \citep{carbonetto2012scalable,carbonetto2017varbvs}. Denoting all unobserved model components by $\theta =(b,s, \gamma, \pi, \tau)$, we approximate the posterior $p(\theta|\bfX,y)$ by a distribution $q(\theta)$ from a restricted class of distributions $\mathcal{Q}$, where the goodness of the approximation is measured in terms of the Kullback-Leibler (KL) divergence, i.e.
\begin{equation}
q \in \arg\min_{q \in \mathcal{Q}} D_{\text{KL}}(q\,||\,p(\theta|\bfX,y)).
\end{equation}
A common and useful choice for distributions in class $\mathcal{Q}$ is the mean-field approximation, i.e. that the distribution factorizes in its parameters. We consider
\begin{align}
q(\theta) = q(b,s,\gamma, \pi, \tau)= \prod_{j=1}^p q(b_j,s_j)q(\gamma)q(\pi)q(\tau),
\label{eq:MFA}
\end{align}
where $b_i$ and $s_i$ are not factorised due to their strong dependencies \citep{titsias2011spike}.

The variational approach leads to an iterative inference algorithm \citep{blei2017variational} by observing that minimizing the KL-divergence is equivalent to maximizing the evidence lower bound $\mathcal{L}$ defined by
\begin{align}
\log(p(y))=\mathcal{L}(q)+D_{\text{KL}}(q \,||\,p(\theta\,|\,\bfX,y)).
\label{eq:ELBOdef}
\end{align}
From this, we have \begin{align}
\mathcal{L}(q)&= \int \log \frac{p(y, \theta)}{q(\theta)}\, q(\theta) \,d\theta\\
&= \int \log p(y, \theta)\, q(\theta)\, d\theta +H(q(\theta)),
\end{align}
with $H(q)=\int - q(\theta)\, \log q(\theta) \, d\theta$ denoting the differential entropy.

Variational methods are based on maximisation of the functional $\mathcal{L}$ with respect to $q$ in order to obtain a tight lower bound on the log model evidence and minimize the KL-distance between the density $q$ and the true (intractable) posterior.
Under a mean-field assumption $q(\theta) = \prod_j q(\theta_j)$, the optimal $q_j$ keeping all other factors fixed is given by
\begin{equation}
\log(q_j^*)(\theta_j)= \mathbb{E}_{-j}(\log(p(y,\theta))) -\text{const}.
\end{equation}
Iterative optimization of each factor results in Algorithm \ref{algo}. Details on the variational inference and the updates can be found in Appendix \ref{appendix1} and \ref{appendix2}. The method is implemented in the freely available R package graper.
From the obtained approximation $q$ of the posterior distribution, we obtain point estimates for the model parameters. In particular, we will use the posterior means $\hat{\beta} = \int \beta \, q(\beta) \, d\beta$, $\hat{\gamma} = \int \gamma \, q(\gamma)\, d\gamma$ and $\hat{\pi} = \int \pi\, q(\pi) \,d\pi$. 

\begin{algorithm}
	\begin{algorithmic}[1]
		\State Input: $\bfX,y, \bigsqcup_{k=1}^G \mathcal{G}_k= \{1,\dots, p\}$
		\State Initialize $\EW s_j=1$, $\EW{\beta_j}$ sampled from $\text{N}(0,1)$,
		$\EW\tau=\EW\gamma_k =1$
		\While{$\mathcal{L}(q)$ has not converged}:		
		\For{$k=1,\dots G$}
		\State Set $q(\pi_k) = \operatorname{Beta}(\pi_k | \alpha^{\pi}_k, \beta^{\pi}_k)$ with
		\begin{align*}
		\alpha^{\pi}_k =d_\pi +\sum_{j\in\mathcal{G}_k} \mathbb{E} s_j \; \text{and} \;
		\beta^{\pi}_k = r_\pi +\sum_{j\in\mathcal{G}_k} (1 - \mathbb{E} s_j)
		\end{align*}
		\vspace{-0.3cm}
		\EndFor
		\For{$j=1,\dots p$}
		\State Set $q(s_j)=\operatorname{Ber}(s_j | \psi_j)$,  $q(b_j|s_j=1) = \text{N}(b_j |\mu_j,\sigma_j^2)$ and \par
		\hskip\algorithmicindent 	$q(b_j|s_j=0) = \text{N}(b_j | 0,(\EW\gamma_{g(j)})^{-1})$ with
		\vspace{-0.2cm}\begin{align*}
		\sigma_j^{2} &= (\mathbb{E} \tau ||\bfX_{\cdot,j}||_2^2 +\mathbb{E}\gamma_{g(j)})^{-1}\\
		\mu_j &= \	\sigma_j^2 \mathbb{E} \tau  \left( -\sum_{i=1}^n \sum_{l\neq j}^p \bfX_{ij} \bfX_{il} \mathbb{E} (\beta_l) + \bfX_{\cdot,j}^T y \right)\\
		\qquad\operatorname{logit}(\psi_j) &= \mathbb{E}\log\frac{\pi_{g(j)}}{1-\pi_{g(j)}} +\frac{1}{2} \log(\mathbb{E}\gamma_{g(j)}) +\frac{1}{2} \log(\sigma_j^{2}) +
		\frac{1}{2}\frac{\mu_j^2}{\sigma_j^{2}}
		\end{align*}
		\EndFor
		\State 	Set $q(\tau) = \Gamma(\tau | \alpha^{\tau}, \beta^{\tau})$ with
			\vspace{-0.2cm}
		\begin{align*}
		\alpha^{\tau} = r_\tau +\frac{n}{2}\text{ and } 
		\beta^{\tau} = d_\tau+\frac{1}{2}\mathbb{E}||y-\bfX\beta||^2_2
		\end{align*}
		\vspace{-0.3cm}
		\For{$k=1,\dots G$}
		\State Set $q(\gamma_k) = \Gamma(\gamma_k | \alpha^{\gamma}_k, \beta^{\gamma}_k)$ with
		\vspace{-0.2cm} 
		\begin{align*}
		\alpha^{\gamma}_k = r_\gamma +\frac{1}{2}|\mathcal{G}_k| \text{ and }  
		\beta^{\gamma}_k = d_\gamma+\frac{1}{2}\sum_{j\in\mathcal{G}_k}\mathbb{E}b_j^2
		\end{align*}
		\vspace{-0.3cm}
		\EndFor
		\State Calculate $\mathcal{L}(q) =\mathbb{E} \log p(y, b, s, \gamma, \pi, \tau) +  H(q) $
		\EndWhile
	\end{algorithmic}
	\caption{Inference algorithm}\label{euclid}
	\label{algo}
	\hrule
	\vspace{0.1cm}
	\textbf{Notes:} \textit{The expectations are taken under the current variational distribution $q$, and $H(q)=\int - q(\theta) \log q(\theta) d\theta$ denotes the differential entropy. We use $\mathcal{F}(x|a)$ to denote the probability density function in $x$ of a distribution $\mathcal{F}$ with parameters $a$, e.g. $\text{\normalfont{Beta}}(x|\alpha,\beta)$. In step 7 it is important to keep track of $v=\bfX \, \mathbb{E} \beta$ in the implementation to obtain linear computational complexity in $p$. We set $r_\tau= r_\gamma= d_\tau= d_\gamma=0.001$ and $d_\pi=r_\pi=1$.}									
\end{algorithm}

\paragraph{Remark on the choice of the mean-field assumption}
An interesting deviation from the standard fully factorized mean-field assumption in Equation (\ref{eq:MFA}) is taking a multivariate variational distribution for the model coefficients. This is easily possible for the dense model ($\pi=1, s=1, b = \beta$), where we can consider the factorization
	\begin{align*}
	q(\beta, \gamma, \tau)=q(\beta)q(\gamma)q(\tau).
	\end{align*}
	In particular, a multivariate distribution is kept for the model coefficients $\beta$ instead of factorizing $q(\beta) = \prod_i q(\beta_i)$. Thereby, this approach allows to capture dependencies between model coefficients in the inferred posterior and is less approximative. We will show below that this can improve the prediction results. However, a drawback of this approach is its computational complexity, as it requires the calculation and inversion of a $p\times p$ covariance matrix in each step. While this can be reduced to a quadratic complexity as described in Appendix \ref{appendix2lin}, this is still prohibitive for many applications. Therefore, we concentrate in the following on the fully factorized mean-field assumption but include comparisons to the multivariate approach in the Results.

\subsection{Extension to logistic regression}
\label{logistic}
The model of Section \ref{modeldef} can be flexibly adapted to other types of generalised linear regression setups with suitable link functions and likelihoods. However, the inference framework needs to be adapted due to loss of conjugacy. Here, we extend the model to the framework of logistic regression with a binary response variable, where we assume that the response follows a Bernoulli likelihood with a logistic link function
\begin{align}
y_i|\beta \sim \text{Ber}(\sigma(x_i^T\beta)) \quad \text{with} \quad \sigma(z)=\frac{1}{1+\exp(-z)}.
\end{align}
While the prior structure and core of the variational inference are identical to the case of a linear model, additional approximations are necessary. For this purpose we adopt \citep{Jaakkola} and approximate the likelihood using a lower bound on the logistic function. For an arbitrary $\xi \in \R$ we have
\begin{equation}
	\sigma(z)\geq \sigma(\xi)\exp\left(\frac{1}{2}(z-\xi)-\eta(\xi)(z^2-\xi^2)\right)
\end{equation}
with $\eta(\xi)=\frac{1}{2\xi}\left(\sigma(\xi)-\frac{1}{2}\right)$.
With this, $\log p(y|\beta)=\sum_{i=1}^n \log(\sigma((2y_i - 1) x_i^T\beta))$ can be bounded by 
\begin{align}
	\begin{split}
\log p(y|\beta) &\geq \frac{1}{2}\sum_i (2y_i-1) x_i^T \beta -\sum_i \eta(\xi_i) (x_i^T \beta)^2   \\
& \qquad+\sum_i \left( \log(\sigma(\xi_i))-\frac{1}{2}\xi_i +\eta(\xi_i) \xi_i^2 \right).
\end{split}
\end{align}
As this approximation restores a quadratic form in $\beta$, the remaining updates can be adopted from the case of a linear model above with the additional variational parameter $\xi$ (see Appendix \ref{appendix2log} for details).

\section{Results}\label{results}
\subsection{Results on simulated data}
First, we evaluated the method on simulated data to test its ability to recover the model coefficients and hyper-parameters per group. For this, a random $\bfX$ matrix was generated from a multivariate normal distribution with mean zero and a Toeplitz covariance structure $\Sigma_{ij}=\rho^{|i-j|}$, and the response was simulated from a linear model with normal error. The $p$ predictors were split into $G=6$ groups of equal size, and the coefficients were simulated from the model as described in Equations (\ref{ss-normal}) and (\ref{ss-ber}) with fixed $\pi_k$ and $\gamma_k$ for each group. In particular, we set $\gamma_k =0.01$ for $k=1,2$, $\gamma_k =1$ for $k=3,4$ and $\gamma_k =100$ for $k=5,6$. For each pair of groups with same $\gamma$-value the sparsity level $\pi_k$ was varied between $\nu$ and $\min(1,1.5\nu)$ for a certain value of $\nu$ determining the sparsity level from 0 (sparse) to 1 (dense). We then varied the number of features $p$, the number of samples $n$, the correlation strength $\rho$, the noise precision $\tau$ and the sparsity level $\nu$ (Table \ref{tab:SimParams}) and generated for each setting ten independent data sets. We evaluated the recovery of the hyper-parameter $\gamma$ and $\pi$ for each group and compared the predictive performance and computational complexity to those of related methods including ridge regression \citep{hoerl1970ridge}, Lasso \citep{tibshirani1996regression}, elastic net \citep{zou2005regularization}, adaptive Lasso \citep{zou2006adaptive}, sparse group Lasso, group Lasso \citep{friedman2010note}, GRridge \citep{wiel2016better}, varbvs \citep{carbonetto2017varbvs} and IPF-Lasso \citep{boulesteix2017ipf}.

\begin{table}[H]
	\small
	\centering
\begin{tabular}{ |l | l |l | l | l |  }
	$p$ &$n$ & $\rho$ & $\tau$ &$\nu$ \\
	\toprule
	60,120,\dots,1200&100 & 0 & 1 &0.2 \\
	300 & 20,40,\dots,500 &0 &1& 0.2\\
	300 & 100 &  0, 0.1,\dots,0.9 &1& 0.2\\
	300 & 100 & 0 &0.01,0.1,\dots,100& 0.2\\
	300 & 100 & 0 &1& 0.001,0.01,\\  &&&& \quad  0.05,\dots,1 
\end{tabular}
	\caption{ \label{tab:SimParams} Simulation parameters: $p$ denotes the number of features, $n$ the number of samples, $\rho$ the correlation strength in $\bfX$, $\tau$ the noise precision and $\nu$ the sparsity level.} 
\end{table}

\subsubsection{Recovery of hyper-parameters}
\begin{figure}
\centering
\includegraphics[width=\linewidth]{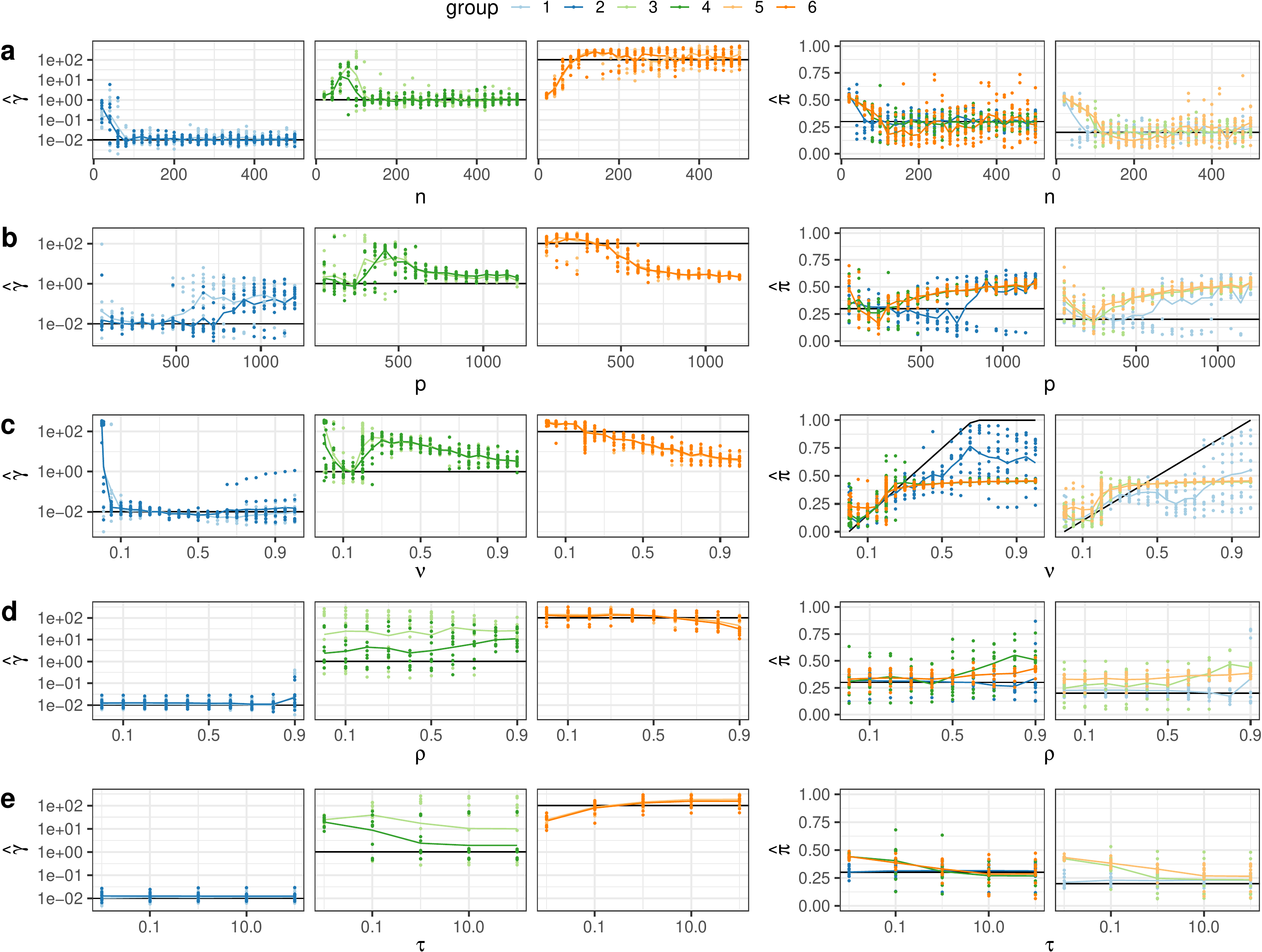}
\caption{Estimated values of the hyperparameter $\gamma$ (left column) and $\pi$ (right column) when varying each of the model parameters (a-e) while keeping the other four parameters fixed as described in Table \ref{tab:SimParams}. The line denotes the mean recovered hyperparameter across 10 random instances of simulated data, while points represent single instances. Colours denote the different groups ($k=1,\dots,6$) and the black line indicates the true value of $\gamma$ (left) and $\pi$ (right) used in the simulation. Each panel displays	 groups with the same value of $\gamma$ (left) and $\pi$ (right).}
\label{fig:sim_hyperparam}
\end{figure}

The algorithm accurately recovered the relative importance of different groups (encoded by $\gamma_k$) and the group-wise sparsity level (encoded by $\pi_k$) across a large range of settings as shown in Figure \ref{fig:sim_hyperparam}. The method failed to recover those parameters accurately only if the ratio between sample size and number of features was too small or the sparsity parameter $\nu$ was too close to 1. These settings were challenging for all methods as can be seen in Section \ref{sec:pred_sim}, where we evaluated estimation and prediction performance in comparison to other methods. In addition, the groups had to contain sufficiently many predictors to reliably estimate group-wise parameters, as seen in Figure \ref{fig:sim_hyperparam}(b). We also noted that a low signal-to-noise ratio could impede the estimation of hyperparameters as can be seen from the group with a very large $\gamma$ value (meaning low coefficient amplitudes as in group 5 and 6) and low precision values ($\tau$) of the noise term.

\subsubsection{Prediction and estimation performance}
\label{sec:pred_sim}
\begin{figure}
	\centering
	\includegraphics[width=\linewidth]{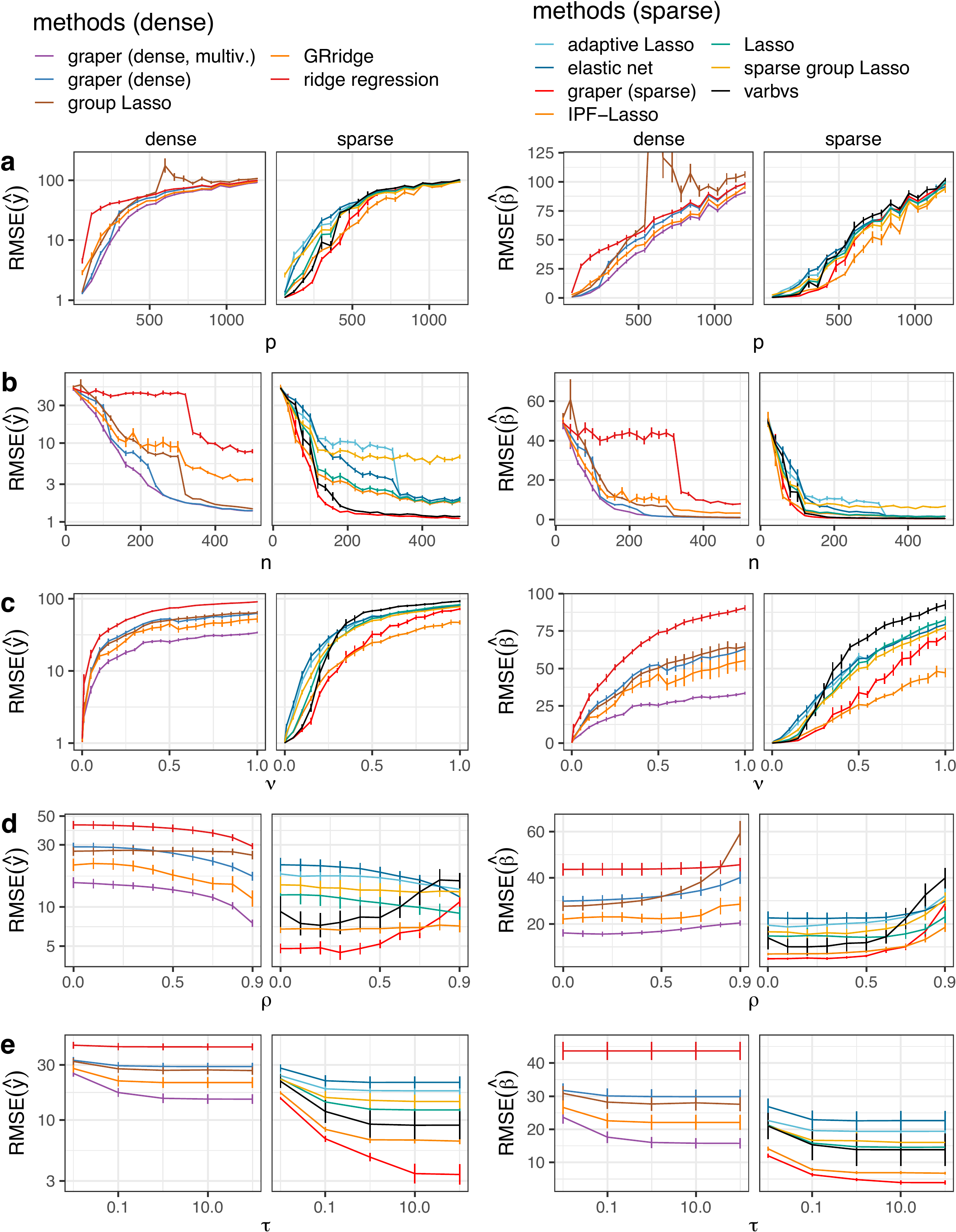}
	\caption{Root mean squared error (RMSE) of the predicted response $\hat{Y} = X^T\hat{\beta}$ (left) and the estimate $\hat{\beta}$ (right) for different methods when varying one of the simulation parameters (a-e) as described in Table \ref{tab:SimParams}. The prediction error is assessed on $n=1000$ test samples. The line denotes the mean RMSE across 10 random instances of simulated data with bars denoting standard errors. The two panels separate methods with sparse estimates of $\beta$ (right) from non-sparse methods (left). (Group Lasso is counted as non-sparse method as it is not sparse within groups.)}
	\label{fig:sim_RMSE_joint}
\end{figure}

Next, we compared the estimation of the true model coefficients and the prediction accuracy on an independent test set of $n=1000$.
Overall, the method showed improved performance for a large range of sample sizes, correlations, numbers of features, noise variances and active features, both in terms of the root mean squared error on $y$ as well as for estimation of $\beta$ (Figure \ref{fig:sim_RMSE_joint}). Among the non-sparse methods {graper} with a non-factorized mean-field assumption clearly outperformed the factorized mean-field assumption as well as GRridge and group Lasso. The covariate-agnostic ridge regression performed worst in most cases. Sparse methods performed in general better in this simulation example, as the underlying model had a large fraction of zero coefficients. Here, we observed that {graper} was comparable to IPF-Lasso, which is the most closely related method. Only in settings with a very high number of active predictors or strong correlations  between the predictors ($\rho$ close to one) the method was outperformed by the IPF-Lasso.

\subsubsection{Scalability}
\begin{figure}
	\centering
	\includegraphics[width=\linewidth]{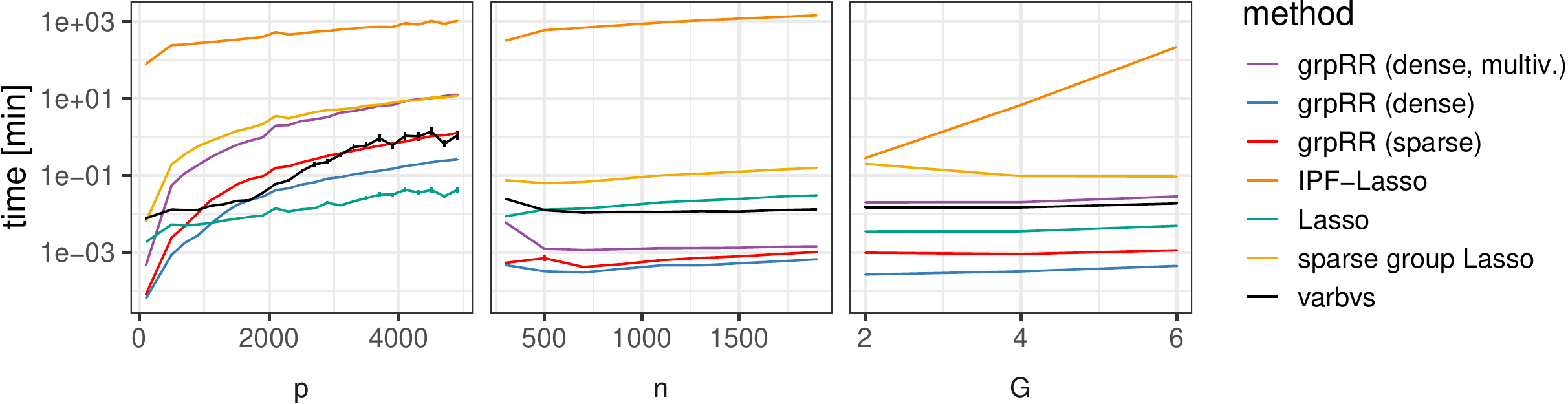}
	\caption{Average run time (in minutes) for different methods when varying the number of samples $n$, features $p$ and groups $G$. Each parameter is varied at a time while holding the others fixed to $n=100$, $p=300$ or $G=6$. Shown are the average times across 50 random instances of simulated data with error bars denoting one standard error.}
	\label{fig:scalability}
\end{figure}

While the additional group-wise optimization comes at a  computational cost, the variational approach runs inference in time complexity linear in the number of features $p$, samples $n$ and groups $G$. Only in the case of a multivariate variational distribution, the complexity is quadratic in the larger of $n$ and $p$ and cubic in the smaller of the two. When varying the number of samples $n$, features $p$ and groups $G$ we observed comparable run times as for Lasso (Figure \ref{fig:scalability}). Differences were mainly observed for $p$: For larger $p$, graper required slightly longer times than Lasso. This difference was more pronounced when using a sparsity promoting spike and slab prior, where additional parameters need to be inferred. As expected, the multivariate approach of graper became considerably slower for large $p$ and showed comparable run times to the sparse group Lasso. The number of groups mainly influenced the computation times of IPF-Lasso, which scales exponentially in the number of groups. Here, graper provided a by far more scalable approach.

\subsection{Application to data from high-throughput biology}

\subsubsection{Drug response prediction in leukaemia samples}
Next, we exemplify the method's performance on real data by considering an application to biological data, where predictors were obtained from different assays. Using assay type as external covariates we used the method to integrate data from the different assays (also referred to as omic types) in a study on chronic lymphocytic leukaemia (CLL)  \citep{cllpaper}. This study combined drug response measurements with molecular profiling including gene expression and methylation. Briefly, we used normalized RNA-Seq expression values of the 5000 most variable genes, the DNA methylation M-values at the 1$\%$ most variable CpG sites as well as the ex-vivo cell viability after exposure to 61 drugs at 5 different concentrations as predictors for the response to a drug (Ibrutinib) that was not included into the set of predictors. In total, this resulted in a model with $n=121$ patient samples and $p=9,553$ predictors. 

\begin{figure}
	\includegraphics[width=1\linewidth]{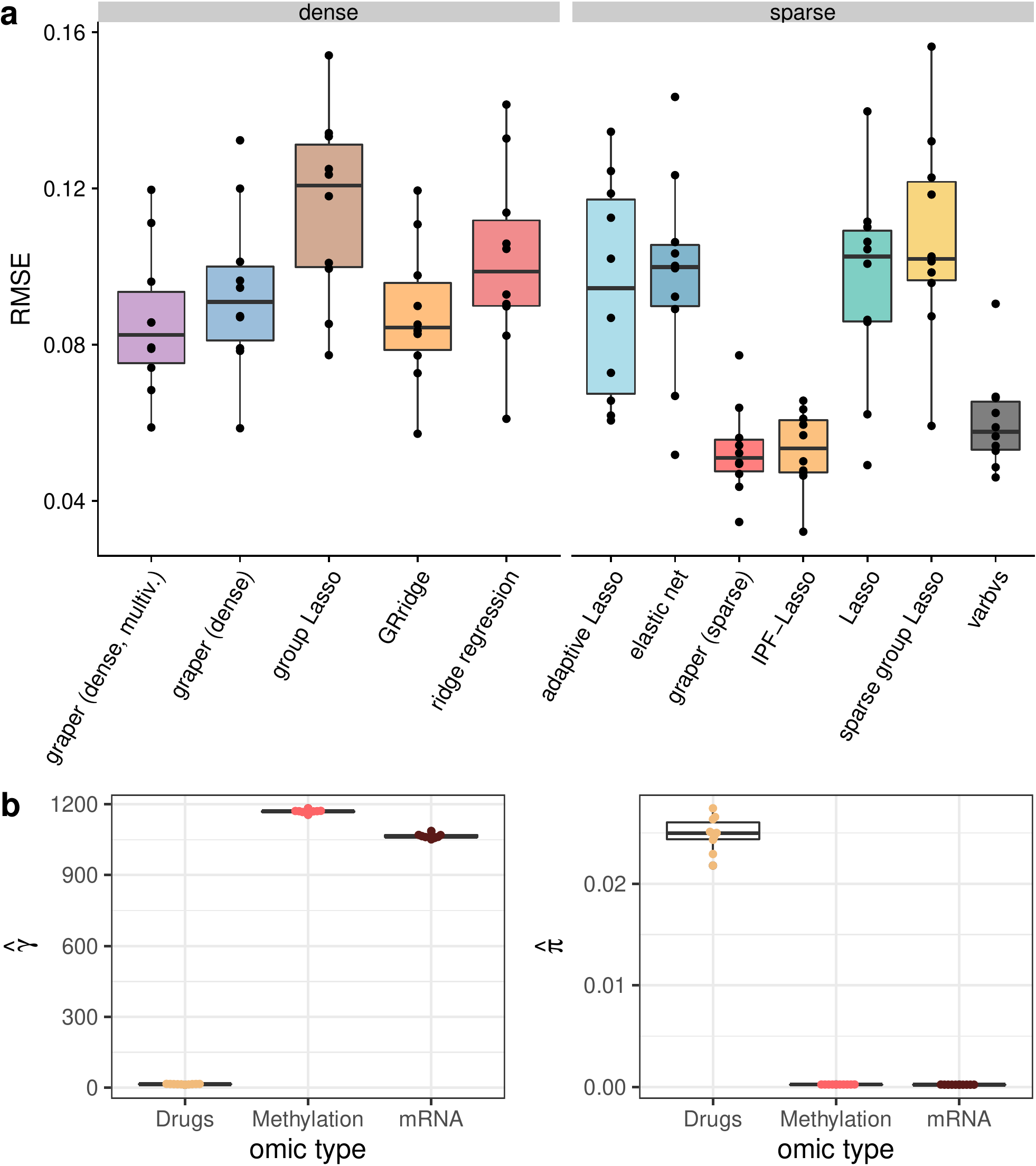}
	\caption{Application to the CLL data with scale differences between assays. \textbf{(a)} Comparison of the root mean-squared error (RMSE) for the prediction of samples' viability after treatment with Ibrutinib. Performance was evaluated in a 10-fold cross-validation scheme, the points denote the individual RMSE for each fold. \textbf{(b)}  Inferred hyperparameters in the different folds for the three different omic types ($\gamma$ on the left and $\pi$ on the right).}
		\label{fig:CLL_unstd}
\end{figure}

We first applied the different regression methods to the data on their original scale. Since the features have different scales (e.g., the drug responses vary from around 1 (neutral) to 0 (completely toxic), the normalized expression values from 0 to 20 and the methylation M-values from -10 to 8),  this ensures that the omic type information is an informative covariate: It results in larger effect sizes of the drug response data and smaller effect sizes of the methylation and expression data compared to scaled predictors. In this setting, incorporating knowledge on the assay type into the penalized regression showed clear advantages in terms of prediction performance: The covariate-aware methods (GRridge, IPF-Lasso and {graper}) all improved upon the covariate-agnostic Lasso, ridge regression or elastic net (Figure \ref{fig:CLL_unstd}(a)). Also the group Lasso methods, which incorporate the group information but apply a single penalty parameter, could not adapt to the scale differences. The inferred hyper-parameters $\gamma$ of {graper} highlighted the larger effect sizes of the drug response feature group, which was strongly favoured by the penalization (Figure \ref{fig:CLL_unstd}(b)). 

\begin{figure}
	\includegraphics[width=1\linewidth]{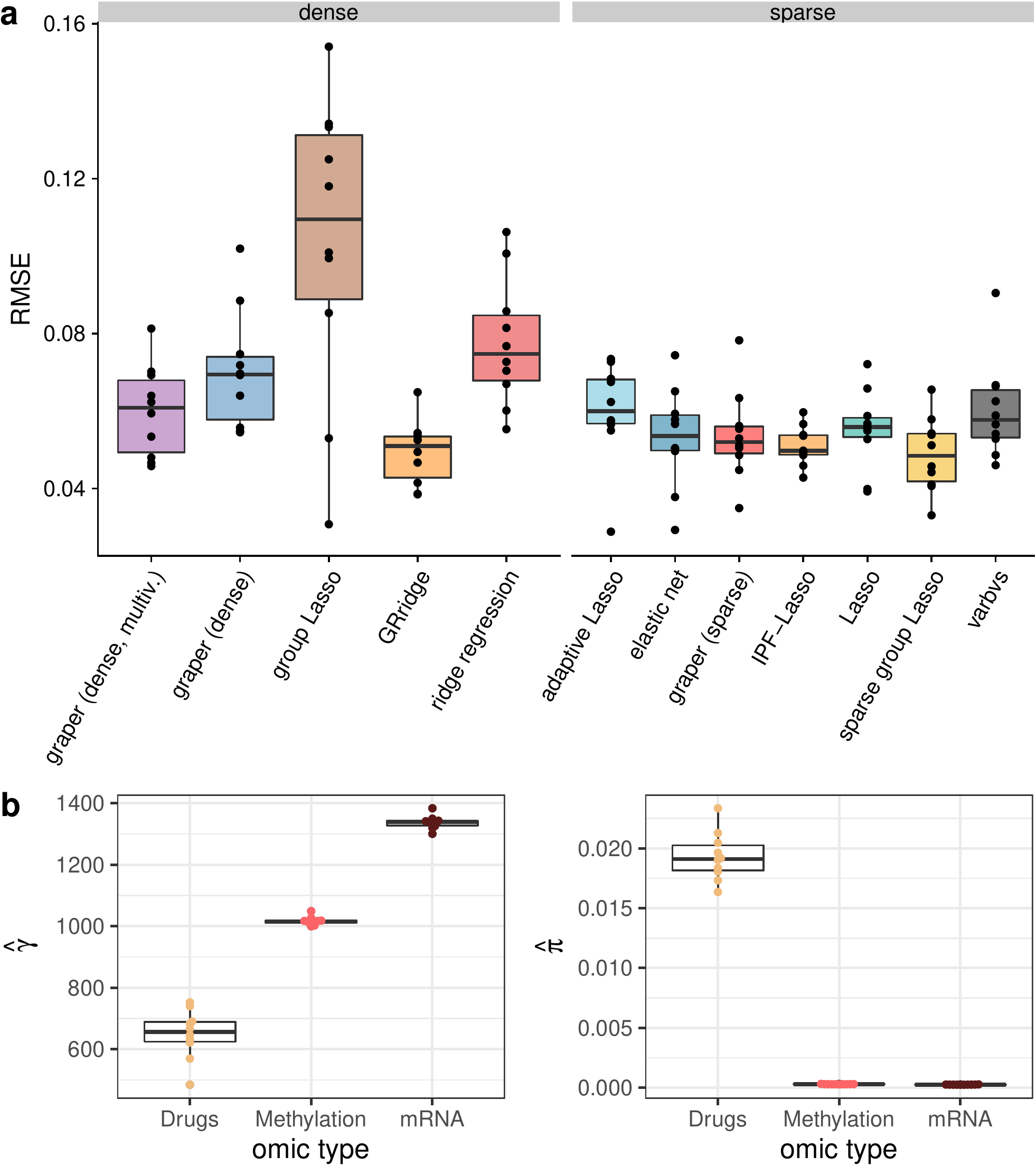}
	\caption{Application to the CLL data with standardized predictors. \textbf{(a)} Comparison of the root mean-squared error (RMSE) for the prediction of samples' viability after treatment with Ibrutinib. Performance was evaluated in a 10-fold cross-validation scheme, the points denote the individual RMSE for each fold. \textbf{(b)} Inferred hyperparameters by graper (sparse) in the different folds for the three different omic types ($\gamma$ on the left and $\pi$ on the right).}
	\label{fig:CLL_std}
\end{figure}

To address differences in feature scale, a common choice made by many implementations (e.g., glmnet \citep{friedman2010regularization}) is to scale all features to unit variance. Indeed, for the data at hand, this transformation was particularly beneficial for the covariate-agnostic methods, and their prediction performances became more similar to those of the covariate-aware methods. However, for dense methods such as ridge regression the covariate information on the omic type remained important (Figure \ref{fig:CLL_std}(a)). Sparse methods in general resulted in very good predictions as the response to Ibrutinib can be well explained by a very sparse model containing only few drugs with related mode of action. By learning weights for each omic type graper directly highlighted the importance of the drug data as predictors (Figure \ref{fig:CLL_std}(b)). 

In general, standardization of all features is unlikely to be an optimal choice, since in many applications there is a relation between information content and amplitude. Here, standardization would drown informative high-amplitude features and 'blow up' noisy low-amplitude features (see Appendix \ref{appendix_std}).

\subsubsection{Age prediction from multi-tissue gene expression data}
As a second example for a covariate in genomics we considered the tissue type. Using data from the GTEx consortium \citep{gtex2013} we asked whether the tissue type is an informative covariate in the prediction of a person's age from gene expression. Briefly, we chose five tissues that were available for the largest number of donors and from each tissue considered the top 50 principal components on the RNA-Seq data after normalization and variance stabilization \citep{love2014moderated}. In total, this gave us $p=250$ predictors from $G=5$ tissues for $n=251$ donors. 

\begin{figure}
		\includegraphics[width=1\linewidth]{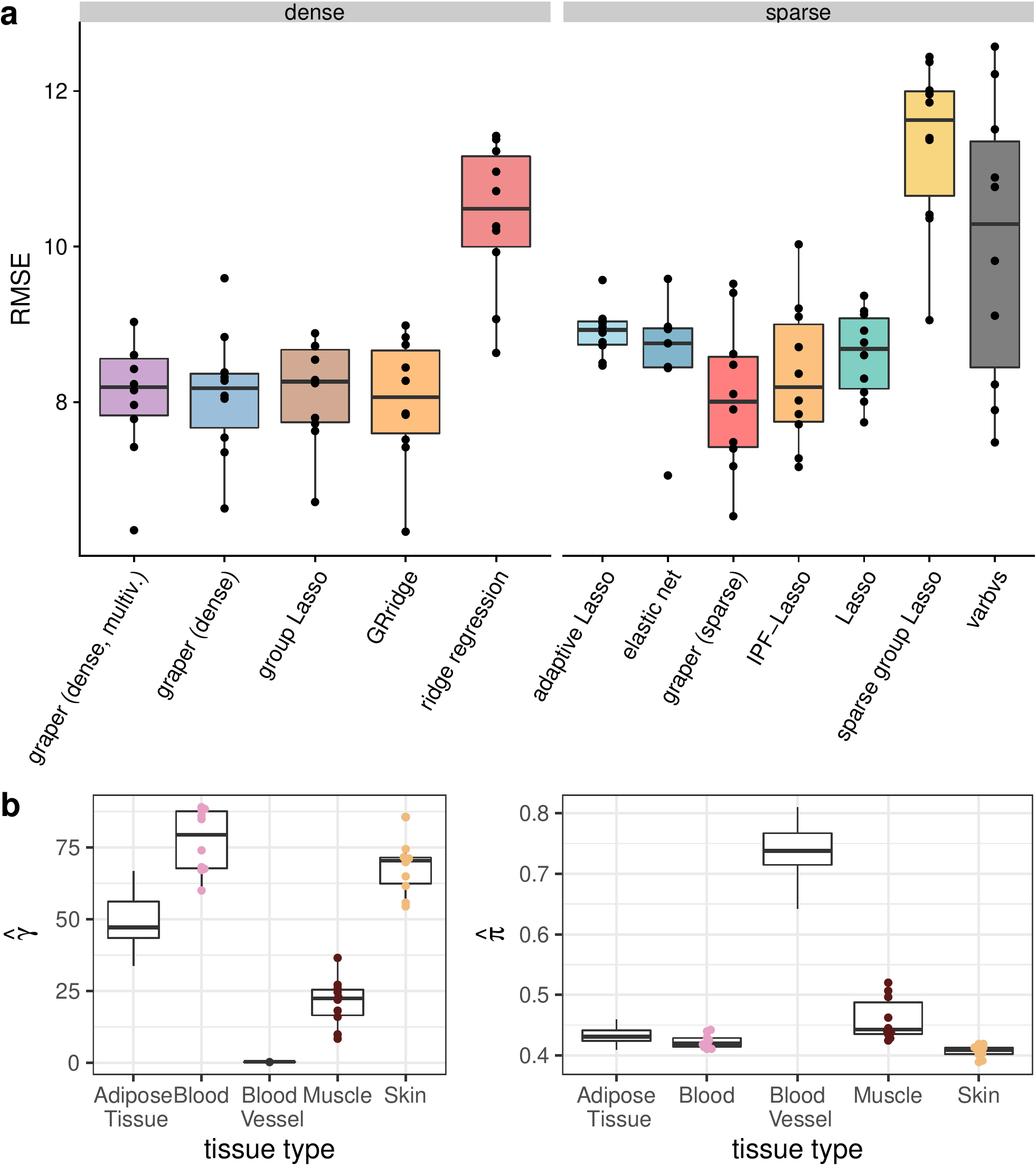}
	\caption{Application to the GTEx data. \textbf{(a)} Comparison of root mean-squared error (RMSE) for the prediction of donor age (in years). Performance is evaluated in a 10-fold cross-validation scheme, the points denote the individual RMSE for each fold. \textbf{(b)} Inferred penalty parameters for the five tissues in graper in each fold. }
	\label{fig:GTEx}
\end{figure}

We observed a small advantage for methods that incorporate the tissue type as a covariate (Figure \ref{fig:GTEx}(a)): GRridge, IPF-Lasso and {graper} all had a smaller prediction error compared to covariate-agnostic methods. In particular, graper resulted in comparable prediction performance to IPF-Lasso whilst requiring less than a second for training compared to 40 minutes for IPF-Lasso. The learnt relative penalization strength and sparsity levels of graper can again provide insights into the relative importance of the different tissue types. In particular, we found lower penalization for blood vessel and muscle and higher penalization for blood and skin (Figure \ref{fig:GTEx}(b)). This is consistent with previous studies on a per-tissue basis, where gene expression in blood vessel has been found to be a good predictor for age, while blood was found to be less predictive \citep{yang2015synchronized}. 

\section{Discussion}

We propose a method that can use information from external covariates to guide penalization in regression tasks and that can provide a flexible and scalable alternative to approaches that were proposed recently \citep{wiel2016better,boulesteix2017ipf}. We illustrated in simulations and data from biological applications that if the covariate is informative of the effect sizes in the model, these approaches can improve upon commonly used penalized regression methods that are agnostic to such information. We investigated the use of important covariates in genomics such as omic type or tissue. The performance of our approach is in many cases comparable to the IPF-Lasso method \citep{boulesteix2017ipf}, while scalability is highly improved in terms of the number of feature groups, thereby extending the range of possible applications.

The variational inference framework provides improved scalability compared to Bayesian methods that are based on sampling strategies. Variational Bayes methods have already been employed in the setting of Bayesian regression with Spike-and-Slab priors \citep{carbonetto2012scalable,carbonetto2017varbvs}. However, these methods do not incorporate information from external covariates. A drawback of variational methods are too concentrated approximations to the posterior distribution. Nevertheless, they have been shown to provide reasonable point estimates in regression tasks \citep{carbonetto2012scalable}, which we focused on here. Due to the mean-field assumption strong correlations between active predictors can lead to suboptimal results of graper. Here, a multivariate mean-field assumption in the variational Bayes approach can be of advantage, suggested as an alternative above. However, it comes at the price of higher computational costs. What is not addressed in our current implementation is the common problem of missing values in the data; if extant, they would need to be imputed beforehand. 

While our approach is related to methods that adapt the penalty function in order to incorporate structural knowledge, such as the group Lasso \citep{yuan2006model}, sparse group Lasso  \citep{friedman2010note} or fused Lasso \citep{ tibshirani2005sparsity}, these approaches apply the same penalty parameter to all the different groups and perform hard in- or exclusion of groups instead of the softer weighting proposed here. Alternatively, the loss function can be modified to incorporate prior knowledge based on a known set of 'high-confidence predictors' as in \citep{jiang2016variable}. The existence and identity of such 'high-confidence predictors', however, is often not clear.

In contrast to frequentist regression methods, the Bayesian approach provides direct posterior-inclusion probabilities for each feature that can be useful for model selection. To obtain frequentist guarantees on the selected features it could be promising to combine the approach with recently developed methods for controlling the false discovery rate (FDR), such as the knockoffs \citep{candes2016panning}. For this, feature statistics can be constructed based on the estimated coefficients or inclusion probabilities from our model as long as the knockoffs obtain the same covariate information as their true counterpart.

An interesting question that we have not addressed is the quest for rigorous criteria when the inclusion of a covariate by differential penalization is of advantage. This question is not limited to the framework of penalised regression but affects the  general setting of shrinkage estimation. While joint shrinkage of a set of estimates can be very powerful in producing more stable estimates with reduced variance, care needs to be taken on which measurements to combine in such a shrinkage approach. As in the case of coefficients in the linear model setting, external covariates could be helpful for this decision and facilitate a more informed shrinkage. However, allowing for differential shrinkage will re-introduce some degrees of freedom into the model and can only be advantageous if the covariate provides 'sufficient' information to balance this. For future work, it would be of interest to find general conditions for when this is the case, thereby enabling an informed choice of covariates in practice.

We provide an open-source implementation of our method in the R package graper. In addition, vignettes and scripts are made available that facilitate the comparison of graper with various related regression methods and can be used to reproduce all results contained in this work.

\newpage
\paragraph{Supplementary Materials}
In Appendix \ref{appendix1} of the Supplementary Materials we provide details on variational inference. Details on the update equations for the proposed methods can be found in Appendix \ref{appendix2}. In Appendix \ref{appendix3} of the Supplementary Materials we discuss practical considerations for the training.

\paragraph{Acknowledgements}
We thank Bernd Klaus for providing useful comments on the manuscript.

\newpage
\bibliography{bibliography}

\newpage
\appendix
\section{Appendix}
Here we provide details on the variational inference scheme, the updates in our model and practical considerations for training. As before, $\bfX$ denotes the $n\times p$ matrix of observed predictors and $y$ the $n$-vector of observed response values. With $\bfX_{ij}$ we denote the $(i,j)$-th element of the matrix $\bfX$ and $\bfX_{\cdot, j}$ its $j$-th column. Furthermore, $y_i \in \R$ denotes the $i$-th response value and $x_i \in \R^p$ the $i$-th predictor vector, corresponding to the $i$-th row in $\bfX$. We will use $\EW= \EW_q$ to denote expectations with respect to the variational distribution $q$. 
\label{appendix}
\subsection{Variational inference}
\label{appendix1}
To arrive at a simple iterative algorithm, we make use of the following lemma, which provides an update rule for each factor in the variational distribution \citep{blei2017variational}.
\begin{lemma}
	\label{lemmaVB}
	Under the mean-field assumption and for a fixed $j$ the evidence lower bound defined in Equation (\ref{eq:ELBOdef}) is maximised by
	$$\log(q_j^*(\theta_j)) = \mathbb{E}_{-j}(\log(p(y,\theta))) -\text{\normalfont{const}},$$
	where the expectation is taken under the current variational distribution $\prod_{l\neq j}q(\theta_l)$.
\end{lemma}
This can be easily seen by writing
\begin{align*}
\mathcal{L}(q) &=\mathbb{E}_q \left(\log \frac{p(y, \theta)}{q(\theta)}\right) \\
& = \int q(\theta) (\log p(y, \theta) - \log q(\theta)) d\theta \\
& = \int q_j(\theta_j) \int (\log p(y, \beta, \gamma, \tau) - \log q(\theta_j)) \prod_{i\neq j}q(\theta_i) d\theta_{-j} d\theta_j \\
&\qquad -\int \sum_{i \neq j}\log q(\theta_i)  \prod_{i\neq j}q(\theta_i) \int q(\theta_j) d\theta_j d\theta_{-j}\\
& = \int q_j(\theta_j) (\EW_{-j}(\log p(y, \theta)) - \log q(\theta_j)) d\theta_j - \text{const}\\
& = \int q_j(\theta_j) \log\left(\frac{\exp \EW_{-j}(\log p(y, \theta))}{q(\theta_j)}\right) d\theta_j - \text{const}\\
& = - D_{\text{KL}}(q_j\,||\,\exp \EW_{-j}(\log p(y, \theta))).
\end{align*}

Hence, after normalising, the distribution $q^*_i(\theta_i)$ is given by
\begin{equation*}
q^*_i(\theta_i) = \frac{\exp(\EW_{-j}[\log p(y, \theta)])}{\int \exp(\EW_{-j}[\log p(y, \theta)]) d\theta_j}.
\end{equation*}

\subsection{Update equations for the variational inference}

\label{appendix2}
\subsubsection{Linear regression model}
\label{appendix2lin}
In the linear model we assume that the likelihood is given by
\begin{equation*}
y|\beta, \tau \sim \text{N}\left(\bfX\beta,\frac{1}{\tau}\mathds{1}\right).
\end{equation*}
With the priors as described in Section \ref{modeldef} and again denoting $\beta = sb$ the joint distribution is given by 
\begin{align*}
p(y,b,s,\gamma,\pi,\tau)=p(y|b,s, \tau) p(b,s|\pi,\gamma)p(\gamma)p(\pi)p(\tau).
\end{align*}
Hence,
\begin{align*}
\log p(y,b&,s,\gamma,\pi,\tau) = \text{const} + \frac{n}{2}\log(\tau)-\frac{\tau}{2}||y-\bfX(b\odot s)||^2_2 \\
&+\sum_{j=1}^p \left\{ \log(\pi_{g(j)})s_j+\log(1-\pi_{g(j)})(1-s_j)\right\}\\
&+\sum_{j=1}^p\left\{\frac{1}{2}\log(\gamma_{g(j)})-\frac{\gamma_{g(j)}}{2}b_j^2\right\}\\
&+\sum_{k=1}^G \left\{(r_\gamma-1)  \log(\gamma_k ) -d_\gamma \gamma_k \right\}\\
&+\sum_{k=1}^G \left\{(d_\pi-1)  \log(\pi_k ) +(r_\pi-1)  \log(1-\pi_k ) -\log(B(d_\pi,r_\pi)\right\}\\
&+(r_\tau-1)  \log(\tau ) -d_\tau \tau.
\end{align*}
The dense model without the spike and slab component arises as a special case when dropping $\pi$ and $s$ from the model and setting $\beta=b$.

For the start we will make a full mean-field assumption, i.e.
	\begin{align*}
	q(b,s,\gamma,\pi,\tau)= \prod_{j=1}^p q(b_j,s_j) q(\gamma) q(\pi) q(\tau),
	\end{align*}
	allowing only a joint distribution for  $(b_j,s_j)$ due to their strong dependencies \citep{titsias2011spike}.
	
Denoting with $\theta$ all individual parameter components in the mean-field assumption, the updates are given by 
	\begin{align*}
	\log(q_j(\theta_j))=\mathbb{E}_{-j}\log(p(y,\theta)),
	\end{align*}
as shown above (Lemma \ref{lemmaVB}).
Thanks to conjugacy between the chosen priors and the likelihood these updates maintain the distributional family of $\theta_j$ reducing the inference to updates of their parameters in each step.  Explicitly, this leads to the following updates in step $l$:
	
	\paragraph{Updates for $\beta$ ($b$ and $s$)}
	For $\beta$ one notes
	\begin{align*}
	\log(q&(b_j, s_j)) \\
	= &- \mathbb{E}\frac{\tau}{2} \mathbb{E}_{-j}||y-\bfX(b\odot s)||^2_2 +\mathbb{E}\log\frac{\pi_{g(j)}}{1-\pi_{g(j)}} s_j -\frac{\mathbb{E}\gamma_{g(j)}}{2}b_j^2 + \text{const}\\
=&-\mathbb{E}\frac{\tau}{2} \left(b_j s_j \sum_k\left(-2y_k\bfX_{kj} + 2 \sum _{l\neq j} \bfX_{kl}\bfX_{kj} \mathbb{E}(s_l b_l)\right)
+ s_j b_j^2 \sum_k \bfX_{kj}^2\right)\\
		& \qquad +\mathbb{E}\log\frac{\pi_{g(j)}}{1-\pi_{g(j)}} s_j-\frac{\mathbb{E}\gamma_{g(j)}}{2}b_j^2 + \text{const}.
	\end{align*}
	This can be written as
		$$q(b_j, s_j) = q(s_j=0)q(b_j|s_j=0)+q(s_j=1)q(b_j|s_j=1),$$
		 where 
	\begin{align*}
	b_j|s_j=0&\sim \text{N}(0,(\mathbb{E}\gamma_{g(j)})^{-1}),\\
	b_j|s_j = 1 &\sim \text{N}(\mu_j^{(l)}, \sigma_j^{(l)2}),
	\end{align*}
	with
	\begin{align*}
	\sigma_j^{(l)2} &= (\mathbb{E} \tau ||\bfX_{\cdot,j}||_2^2 +\mathbb{E}\gamma_{g(j)})^{-1},\\
	\mu_j^{(l)} &= \	\sigma_j^{(l)2} \mathbb{E} \tau  \left( -\sum_{k=1}^n \sum_{l\neq j}^p \bfX_{kj} \bfX_{kl} \mathbb{E} (\beta_l) + \bfX_{\cdot,j}^T y \right).
	\end{align*}
	To make this scale linearly in $p$ in the inner loop we follow \citep{carbonetto2012scalable} and keep track of $v=\bfX\mu$ and update this only in the new component $v \leftarrow v + (\mu_j^{(\text{new})} - \mu_j)\bfX_{\cdot,j}$.
	
	The marginal distribution of $s_j$ is given by	$s_j \sim \text{Ber}(\psi_j^{(l)})$ with $\psi_j^{(l)}$ obtained from
	\begin{align*}
	\text{logit}(\psi_j^{(l)})&= \mathbb{E}\log\frac{\pi_{g(j)}}{1-\pi_{g(j)}} -  
	\frac{1}{2}  \log(\mathbb{E} \tau ||\bfX_{\cdot,j}||_2^2 +\mathbb{E}\gamma_{g(j)}) + \frac{1}{2}  \log(\mathbb{E}\gamma_{g(j)}) \\
	& \qquad+\frac{( \mathbb{E} \tau)^2  \left(\bfX_{\cdot,j}^T y -\sum_{k=1}^n \sum_{l\neq j}^p \bfX_{kj} \bfX_{kl} \mathbb{E} (b_l s_l) \right)^2}{2 (\mathbb{E} \tau ||\bfX_{\cdot,j}||_2^2 +\mathbb{E}\gamma_{g(j)})^{-1}}\\
	&= \mathbb{E}\log\frac{\pi_{g(j)}}{1-\pi_{g(j)}} +\frac{1}{2} \log(\mathbb{E}\gamma_{g(j)}) +\frac{1}{2} \log(\sigma_j^{2}) +
		\frac{1}{2}\frac{\mu_j^2}{\sigma_j^{2}}.
	\end{align*}
	This is derived by integrating the joint density of $q(b_j,s_j)$ to obtain the marginal density of $s_j$. Denoting the normal density with $\varphi(\cdot; \mu,\sigma^2)$ we have
	\begin{align*}
	q(s_j)&= \int q(b_j,s_j) db_j \\
	&\propto \exp\left(\mathbb{E}\log\frac{\pi_{g(j)}}{1-\pi_{g(j)}} s_j \right) \\
	& \qquad\int \exp \left(- \mathbb{E}\frac{\tau}{2} \mathbb{E}_{-j}||y-\bfX(b\odot s)||^2_2 -\frac{\mathbb{E}\gamma_{g(j)}}{2}b_j^2\right) db_j \\
	&\propto \exp\left(\mathbb{E}\log\frac{\pi_{g(j)}}{1-\pi_{g(j)}} s_j \right) \\
	& \qquad \int \varphi(b_j; \mu_j(s_j), \sigma_j^2(s_j)) \sqrt{2\pi\sigma_j^2(s_j)} \exp\left( \frac{\mu_j(s_j)^2}{2\sigma_j^2(s_j)} \right) db_j \\
	&\propto \exp\left(\mathbb{E}\log\frac{\pi_{g(j)}}{1-\pi_{g(j)}} s_j \right) \sqrt{\sigma_j^2(s_j)} \exp\left( \frac{\mu_j(s_j)^2}{2\sigma_j^2(s_j)} \right)\cdot 1\\
	&= \exp\left(\mathbb{E}\log\frac{\pi_{g(j)}}{1-\pi_{g(j)}} s_j +\frac{1}{2} \log{\sigma_j^2(s_j)}  +\frac{\mu_j(s_j)^2}{2\sigma_j^2(s_j)} \right).\\
	\intertext{Hence,}
	\log q(s_j)&= \text{const}+ \mathbb{E}\log\frac{\pi_{g(j)}}{1-\pi_{g(j)}} s_j +\frac{1}{2} \log{\sigma_j^2(s_j)} +\frac{\mu_j(s_j)^2}{2\sigma_j^2(s_j)} \\
	&= \text{const}+  s_j  \mathbb{E}\log\frac{\pi_{g(j)}}{1-\pi_{g(j)}} -\frac{1}{2}  \log(s_j\mathbb{E} \tau ||\bfX_{\cdot,j}||_2^2 +\mathbb{E}\gamma_{g(j)})\\
	& \qquad+\frac{s_j^2 ( \mathbb{E} \tau)^2  \left(\bfX_{\cdot,j}^T y -\sum_{k=1}^n \sum_{l\neq j}^p \bfX_{kj} \bfX_{kl} \mathbb{E} (b_l s_l) \right)^2}{2 (s_j\mathbb{E} \tau ||\bfX_{\cdot,j}||_2^2 +\mathbb{E}\gamma_{g(j)})^{-1}}\\
	&=\text{const} +s_j \left\{\mathbb{E}\log\frac{\pi_{g(j)}}{1-\pi_{g(j)}} -  
	\frac{1}{2}  \log(\mathbb{E} \tau ||\bfX_{\cdot,j}||_2^2 +\mathbb{E}\gamma_{g(j)}) \right. \\ 
	& \qquad\left.+\frac{1}{2}  \log(\mathbb{E}\gamma_{g(j)})+\frac{( \mathbb{E} \tau)^2  \left(\bfX_{\cdot,j}^T y -\sum_{k=1}^n \sum_{l\neq j}^p \bfX_{kj} \bfX_{kl} \mathbb{E} (b_l s_l) \right)^2}{2 (\mathbb{E} \tau ||\bfX_{\cdot,j}||_2^2 +\mathbb{E}\gamma_{g(j)})^{-1}} \right\}.
	\end{align*}
	In the last steps note $s\in \{0,1\}$. Comparing this to $s\sim \text{Ber}(\psi)$ where $\log(q(s))=\text{const} + s\, \text{logit}(\psi)$ we get the above formula for $\psi^{(l)}$.

	Taken together, $\beta_j=s_jb_j \sim \delta_0 (1-\psi^{(l)}_j) + \psi^{(l)}_j \text{N}(\mu_j^{(l)},\sigma_j^{(l)2})$.

	\paragraph{Updates for $\gamma= (\gamma_1, \dots, \gamma_G)$}
	\begin{align*}
	\log q(\gamma)&=\text{const}+ 
	\sum_{j=1}^p \left\{ \frac{1}{2}\log(\gamma_{g(j)}) -\frac{\gamma_{g(j)}}{2}\EW b_j^2 \right\}\\
	& \qquad +
	\sum_{k=1}^G \left\{(r_\gamma-1)  \log(\gamma_k ) -d_\gamma \gamma_k \right\}\\
	&=\text{const}+\sum_{k=1}^G \left\{  \log(\gamma_k )  (r_\gamma-1 +\frac{1}{2}|\mathcal{G}_k|) - \gamma_k (d_\gamma+\frac{1}{2}\sum_{j\in\mathcal{G}_k}\EW b_j^2)\right\}
	\end{align*}
	Thus, $\gamma_k \sim \Gamma(\alpha^{\gamma,(l)}_k, \beta^{\gamma,(l)}_k)$ are independent gamma distributions with parameters in step $l$ given by
	\begin{align*}
	\alpha^{\gamma,(l)}_k &= r_\gamma +\frac{1}{2}|\mathcal{G}_k|,\\
	\beta^{\gamma,(l)}_k &= d_\gamma+\frac{1}{2}\sum_{j\in\mathcal{G}_k}\mathbb{E}b_j^2.
	\end{align*}

	\paragraph{Updates for $\tau$}
	\begin{align*}
	\log q(\tau)&= \text{const}+ \frac{n}{2} \log(\tau) -\frac{\tau}{2} \EW||y-\bfX\beta||^2_2
	+(r_\tau-1)  \log(\tau ) -d_\tau \tau 
	\end{align*}
	Thus, $\tau \sim \Gamma(\alpha^{\tau, (l)}, \beta^{\tau,(l)})$ is  a gamma distribution with parameters  in step $l$ given by
	\begin{align*}
	\alpha^{\tau, (l)} &= r_\tau +\frac{n}{2},\\
	 \beta^{\tau,(l)} &= d_\tau+\frac{1}{2}\EW\beta||y-\bfX\beta||^2_2.
	\end{align*}

	\paragraph{Updates for $\pi= (\pi_1, \dots,\pi_G)$}
	\begin{align*}
	\log q(\pi)&= \text{const}+ \sum_{j=1}^p \log(\pi_{g(j)})\EW s_j+\log(1-\pi_{g(j)})(1-\EW s_j)\\
	& \qquad+\sum_{k=1}^G \left\{(d_\pi-1)  \log(\pi_k ) +(r_\pi-1)  \log(1-\pi_k ) -\log(B(d_\pi,r_\pi)\right\}\\
	&=\sum_{k=1}^G \log(\pi_k) (d_\pi-1 +\sum_{j\in\mathcal{G}_k} \EW s_j) +  \log(1-\pi_k )(r_\pi-1 +\sum_{j\in\mathcal{G}_k} 1 - \EW s_j)\\
	\end{align*}
	Thus, $\pi_k \sim \text{Beta}(\alpha^{\pi,(l)}_k, \beta^{\pi,(l)}_k)$ are independent beta distributions with parameters in step $l$ given by
	\begin{align*}
	\alpha^{\pi,(l)}_k &=d_\pi +\sum_{j\in\mathcal{G}_k} \EW s_j,\\
	\beta^{\pi,(l)}_k &= r_\pi +\sum_{j\in\mathcal{G}_k} (1 - \EW s_j).
	\end{align*}

	\paragraph{Expected values required}
	The updates above involve the calculation of expected values under the current variational distribution $q$. These are given by 
	\begin{align*}
	\mathbb{E} \tau &= \frac{\alpha^{\tau}}{\beta^{\tau}},\\
	\mathbb{E}\gamma_{k} &= \frac{\alpha^{\gamma}_k}{\beta^{\gamma}_k},\\
	\mathbb{E}\log\frac{\pi_k}{1-\pi_k} & =  \psi(\alpha^{\pi}_k) - \psi(\beta^{\pi}_k),\\
	\mathbb{E} s_j & = \psi_j,\\
	\mathbb{E}||y-\bfX\beta||^2_2 &=y^T y -2 y^T \bfX \mu_\beta + \sum_{i,j} (\bfX^T\bfX)_{i,j} (\Sigma^\beta_{i,j} +\mu^\beta_{i}\mu^\beta_{j}),   \\
	\mathbb{E}b_j &= \psi_j\mu_j,\\
	\mathbb{E}b_j^2 &= (1-\psi_j) \left(\mathbb{E}\gamma_{g(j)}^{-1}\right) + \psi_j\left(\mu_j^2 + \sigma_j^2\right),\\
	\mathbb{E} \beta_j = \mathbb{E} b_j s_j &=\mu_j \psi_j,\\
	\mathbb{E} \beta_j^2 = \mathbb{E} b_j^2 s_j &=(\mu_j^2 +  \sigma_j^2) \psi_j.
	\end{align*}
	Here, $\psi$ denotes the digamma function $\psi(x)=\frac{\Gamma'(x)}{\Gamma(x)}$ and
	\begin{align*}
\mu^\beta &= (\mathbb{E} \beta_j)_{j=1,\dots,p} = (\mathbb{E} b_j s_j)_{j=1,\dots,p},\\
\Sigma^\beta&= \text{diag}(\Var(\beta_j)_{j=1,\dots,p})= \text{diag}( (\mathbb{E} \beta_j^2 - (\mathbb{E} \beta_j)^2)_{j=1,\dots,p}).
	\end{align*}
	Note that here and in the following we dropped the step index $(l)$ of all parameters from the notation for simplicity.
	
	\paragraph{Calculation of the Evidence Lower Bound}
	The evidence lower bound bounds the log model evidence from below and can be calculated in each step to monitor convergence. Recall
	$$\log(y)=\mathcal{L}(q)+D_{\text{KL}}(q\,||\,p)$$
	with
	
	\begin{align*}
	\mathcal{L}(q)&=\mathbb{E}_q \left(\log \frac{p(y, b, s, \gamma, \pi, \tau)}{q(b, s, \gamma, 
		\pi, \tau)}\right)\\
	&=\mathbb{E}_q \left(\log p(y, b, s, \gamma, 
	\pi, \tau)\right) +H(q(b, s, \gamma, 
	\pi, \tau))\\
	&=\mathbb{E}_q \left(\log p(y, b, s,\gamma, 
	\pi, \tau)\right) + \sum_{j=1}^p H(q(b_j, s_j)) \\
	& \qquad+ H(q(\gamma)) + H(q(\pi)) +H(q(\tau)),
	\end{align*}
	where $H(q)=\int - q(\theta) \log q(\theta) d\theta$ denotes the differential entropy.
The terms from the joint model density are given by
	\begin{align*}
	\mathbb{E}_q \log p(y, b, s, \gamma, \tau)&=\mathbb{E}_q \log p(y| b, s, \tau) +\mathbb{E}_q \log p(b| \gamma)  +\mathbb{E}_q \log p(s| \pi) \\
	& \qquad +\mathbb{E}_q \log p(\gamma) +\mathbb{E}_q \log p(\pi) +\mathbb{E}_q \log p(\tau)
	\end{align*}
	with 
	\begin{align*}
	\mathbb{E}_q \log p(y| \beta, \tau) &= \frac{n}{2} \mathbb{E} \log(\tau) -\frac{1}{2}\mathbb{E}\tau ||y-\bfX(b \odot s)||_2^2-\frac{n}{2} \log(2\pi),\\
	\mathbb{E}_q \log p(b| \gamma) &= \sum_j \left(\frac{1}{2} \mathbb{E} \log(\gamma_{g(j)}) - \frac{1}{2} \mathbb{E} \gamma_{g(j)}b_j^2 - \frac{1}{2} \log(2\pi)\right),\\ 
	\mathbb{E}_q \log p(s| \pi) &= \sum_j \left( \mathbb{E} s_j \log (\pi_{g(j)}) + \mathbb{E} (1-s_j) \log (1-\pi_{g(j)})\right),\\ 
	\mathbb{E}_q \log p(\gamma) &=\sum_k \left((r_\gamma-1)\mathbb{E} \log(\gamma_k) - d_\gamma \mathbb{E} \gamma_k - \log(\Gamma((r_\gamma)) + r_\gamma \log(d_\gamma)\right), \\
	\mathbb{E}_q \log p(\pi) & =\sum_k \left( (d_\pi -1) \EW \log(\pi_k) + (r_\pi -1 ) \EW \log(1- \pi_k) - \log \text{B}(d_\pi, r_\pi)\right),\\
	\mathbb{E}_q \log p(\tau) &=(r_\tau-1)\mathbb{E} \log(\tau) - d_\tau \mathbb{E} \tau - \log(\Gamma((r_\tau)) + r_\gamma \log(d_\tau).
	\end{align*}
	Here, $B(a,b) = \frac{\Gamma(a)\Gamma(b)}{\Gamma(a+b)}$ denoted the beta function.
	The required expectations in addition to those used in the updates are easily obtained using the known distributions and parameters of the variational density in each iteration and the fact that $q$ factorizes, i.e.
	\begin{align*}
	\mathbb{E} \log(\tau)&=\psi(\alpha_\tau) - \log(\beta_\tau),\\
	\mathbb{E}\tau ||y-\bfX(b \odot s)||_2^2&=\mathbb{E}\tau \mathbb{E}||y-\bfX(b \odot s)||_2^2,\\
	\mathbb{E} \log(\gamma_k)&=\psi(\alpha^{\gamma}_k) - \log(\beta^{\gamma}_k)\\
	\mathbb{E} \gamma_{g(j)}\beta_j^2 &=\mathbb{E} \gamma_{g(j)}\mathbb{E}\beta_j^2,\\
	\mathbb{E} \log(\pi_{k}) & = \psi(\alpha^{\pi}_k) - \psi(\alpha^{\pi}_k + \beta^{\pi}_k),\\
		\mathbb{E} (1-\log(\pi_{k})) & = \psi(\beta^{\pi}_k) - \psi(\alpha^{\pi}_k + \beta^{\pi}_k).
	\end{align*}
 The entropies are derived from the known expression for the entropy of the gamma, beta, Bernoulli and normal distribution, i.e.
	\begin{align*}
	H(q(b_j, s_j)) &= H(q(b_j|s_j)) + H(q(s_j))\\
	H(q(b_j|s_j)) &=\frac{1}{2}(\log(2\pi)+1)-\frac{1}{2} \log(s_j\EW\tau ||\bfX_{\cdot,i}||_2^2 + \gamma_{g(j)})\\
	H(q(s_j)) &= - (1-\psi_j) \log(1-\psi_j) - \psi_j \log(\psi_j)\\
	H(q(\gamma)) &=\sum_k \left(\alpha^{\gamma}_k - \log(\beta^{\gamma}_k) + \log (\Gamma(\alpha^{\gamma}_k)) + (1-\alpha^{\gamma}_k)\psi(\alpha^{\gamma}_k) \right)\\
	H(q(\pi)) &=\sum_k  \left(\log \,\text{B}(\alpha^{\pi}_k, \beta^{\pi}_k) - (\alpha^{\pi}_k-1)\psi(\alpha^{\pi}_k)- (\beta^{\pi}_k-1)\psi(\beta^{\pi}_k) \right.\\
	& \left.\qquad +\, (\alpha^{\pi}_k + \beta^{\pi}_k -2) \psi(\alpha^{\pi}_k + \beta^{\pi}_k) \right)\\
	H(q(\tau)) &=\alpha^{\tau} - \log(\beta^{\tau}) + \log (\Gamma(\alpha^{\tau})) + (1-\alpha^{\tau})\psi(\alpha^{\tau}).
	\end{align*}

\paragraph{Multivariate mean-field approximation for $\beta$}
The assumption that the variational distribution $q(\beta)$ factorizes across all predictors can be very strong. Therefore, a more accurate approximation of the true posterior can be obtained by allowing for a $p$-variate distribution for $\beta$.

For $s=1$, i.e. no spike term in the model, and hence $\beta=b$ the joint distribution in the updates is then given by
	\begin{align*}
	\log q(\beta)&=\text{const} -\frac{\mathbb{E}(\tau)}{2}||y-\bfX\beta||^2_2+
	\sum_{j=1}^p \left\{-\frac{\mathbb{E}(\gamma_{g(j)})}{2}\beta_j^2 \right\}
	\end{align*}	
	Thus, $\beta \sim \text{N}(\mu^{(l)}, \Sigma^{(l)})$ is a normal distribution with parameters
	\begin{align*}
	\mu^{(l)} &=\mathbb{E}(\tau)\Sigma^{(l)} \bfX^T y, \\
	\Sigma^{(l)} &= (\mathbb{E}(\tau) \bfX^T \bfX +D)^{-1} \quad \text{with}\, D=\text{diag}((\mathbb{E}(\gamma_{g(j)}))_{j=1,\dots,p}).
	\end{align*}
	The other updates stay the same, where the covariance matrix $\Sigma$ is now no longer diagonal as previously.
	As this update requires the inversion of a $p\times p$ matrix a limiting factor for applying the multivariate mean-field approximation is its computational complexity. When $n$ is small compared to $p$ a better solution is to employ the Woodbury-Matrix identity \citep{woodbury1950inverting}, i.e.
	\begin{equation*}
	\Sigma^{(l)} = D-D\bfX^T((\mathbb{E}(\tau))^{-1} \mathds{1}_n + \bfX D\bfX^T)^{-1}\bfX D,
	\end{equation*}
	which requires the inversion of a $n \times n$ matrix only.
	This multivariate assumption can be useful in the presence of strong correlations between the predictors. In the case where $\bfX^T\bfX$ is diagonal we obtain a similar form than for a fully factorized variational distribution.
	
		The evidence lower bound is obtained analogous to the fully factorized case with a multivariate normal distribution and dropping the terms involving $s$ and $\pi$. In particular
		\begin{align*}
		\mathcal{L}(q)
		&=\mathbb{E}_q \left(\log p(y, \beta,\gamma, \tau)\right) +  H(q(\beta)) + H(q(\gamma))  +H(q(\tau)),
		\end{align*}
		with 
		\begin{align*}
		H(q(\beta)) &=\frac{p}{2}(\log(2\pi)+1)+\frac{1}{2} \log(|\Sigma|).
		\end{align*}

	\subsubsection{Logistic regression model}
	\label{appendix2log}
	In order to adapt the model to binary data, we change the likelihood of $y$ to a Bernoulli distribution and consider a generalized linear model with logistic link function, i.e.
		\begin{align*}
		y_i|\beta &\sim \text{Ber}(\sigma(x_i^T\beta)) \quad \text{with} \quad \sigma(z)=\frac{1}{1+\exp(-z)}.
		\end{align*}
	The priors on the model coefficients $\beta$ remain the same as in the linear model, the noise variance $\tau$ is dropped from the model.
	While the model is strongly related to the case of the normal response variable, the challenge here lies in the fact that with the Bernoulli distribution for $Y$ we loose the conjugacy of the prior from the linear model. To solve this and still obtain a fast and explicit inference scheme, we use an approximation of the sigmoid function by an exponential of a quadratic term, thus restoring conjugacy.
	
		As $\sigma(-a) = 1- \sigma(a)$ we can write
		\begin{align*}
		\mathbb{P}(y_i=1|\beta)&=\sigma(x_i^T\beta), \\
		\mathbb{P}(y_i=0|\beta)&=\sigma(-x_i^T\beta),
		\end{align*}
		and hence the likelihood is given by
		$$ p(y_i|\beta)=\sigma((2y_i-1)x_i^T\beta).$$
		Following \citep{Jaakkola} we use the following lower bound on the sigmoid
		$$\sigma(z)\geq \sigma(\xi)\exp\left(\frac{1}{2}(z-\xi)-\eta(\xi)(z^2-\xi^2)\right), \qquad \eta(\xi)=\frac{1}{2\xi}\left(\sigma(\xi)-\frac{1}{2}\right).$$
		This introduces an additional variational parameter $\xi$, which we update alongside the other updates to improve this approximation in each iteration.

	Using this approximation we can bound the joint density of the model by
			\begin{align*}
			p(y,\beta,\gamma)&=p(y|\beta)p(\beta|\gamma,\pi)p(\gamma)p(\pi)\\
			&\geq h(\beta, \xi) p(\beta|\gamma,\pi)p(\gamma)p(\pi),
			\end{align*}
		with 
		\begin{equation}
		\label{jaakkola_bound}
		\begin{split}
\log h(\beta, \xi)&=\frac{1}{2}\sum_i (2y_i-1) x_i^T \beta -\sum_i \eta(\xi_i) (x_i^T \beta)^2  \\
& \qquad + \sum_i \left( \log(\sigma(\xi_i))-\frac{1}{2}\xi_i +\eta(\xi_i) \xi_i^2 \right).
		\end{split}
		\end{equation}	
	With the fully factorised mean-field assumption we get the following updates:
			\begin{align*}
			\log(q(b_j, s_j)) &=\text{const} +\log h(\beta, \xi) -\frac{\mathbb{E}(\gamma_{g(j)})}{2}b_j^2 +\mathbb{E}\log\frac{\pi_{g(j)}}{1-\pi_{g(j)}} s_j\\
			&= \text{const}+ \frac{1}{2}\sum_i (2y_i-1) x_i^T \beta -\sum_i \eta(\xi_i) (x_i^T \beta)^2\\  &\quad-\frac{\mathbb{E}(\gamma_{g(j)})}{2}b_j^2 +\mathbb{E}\log\frac{\pi_{g(j)}}{1-\pi_{g(j)}} s_j\\
			&= \text{const}+ \left(\sum_i (y_i-\frac{1}{2})\bfX_{ij}\right)  b_j s_j -
			s_jb_j^2 \sum_{i=1}^n \eta(\xi_i) \bfX_{ij}^2 \\
			& \quad-
			2b_j s_j \sum_{i=1}^n \eta(\xi_i) \sum_{l\neq j} \bfX_{il}\bfX_{ij} \mathbb{E} \beta_l -\frac{\mathbb{E}(\gamma_{g(j)})}{2}b_j^2 +\mathbb{E}\log\frac{\pi_{g(j)}}{1-\pi_{g(j)}} s_j.
			\end{align*}
Analogous to the linear model we can derive the following updates for the coefficients: $b_j|s_j=0 \sim \text{N}(0,\mathbb{E}\gamma_{g(j)}^{-1})$ and  $b_j|s_j=1 \sim \text{N}(\mu_j,\sigma_j^2)$ with
		\begin{align*}
		\sigma_j^2&=\left(2 \sum_{i=1}^n \eta(\xi_i) \bfX_{ij}^2 +\mathbb{E}\gamma_{g(j)}\right)^{-1},\\
		\mu_j&=\sigma_j^2 \left(-2 \sum_{i=1}^n \eta(\xi_i) \sum_{l\neq j}^p \bfX_{ij}\bfX_{il} \mathbb{E} \beta_l +  \bfX^T_{\cdot,j}(y-\frac{1}{2})\right).
		\end{align*}
		The probability for $s_j=1$ is given by
		$$\text{logit}(\psi_j^{(l)})= \mathbb{E}\log\frac{\pi_{g(j)}}{1-\pi_{g(j)}} -\frac{1}{2} \log(\mathbb{E}\gamma_{g(j)}^{-1}) +\frac{1}{2} \log(\sigma_j^{2}) +
		\frac{1}{2}\frac{\mu_j^2}{\sigma_j^{2}},$$
			as in the linear model.	

	In the case of a multivariate mean-filed assumption on $\beta$ we obtain
		\begin{align*}
		\log q(\beta)&=\text{const} +\log h(\beta, \xi)+
		\sum_{j=1}^p \left\{-\frac{\mathbb{E}(\gamma_{g(j)})}{2}\beta_j^2 \right\}\\
		&= \text{const}+ \frac{1}{2}\sum_{i=1}^n (2y_i-1) x_i^T \beta -\sum_{i=1}^n \eta(\xi_i) (x_i^T \beta)^2 +
		\sum_{j=1}^p \left\{-\frac{\mathbb{E}(\gamma_{g(j)})}{2}\beta_j^2 \right\}\\
		&= \text{const}+ \left(\sum_{i=1}^n (y_i-\frac{1}{2})x_i^T\right)  \beta -\beta^T \left(\sum_{i=1}^n \eta(\xi_i) x_i x_i^T\right) \beta \\
		& \qquad +
		\sum_{j=1}^p \left\{-\frac{\mathbb{E}(\gamma_{g(j)})}{2}\beta_j^2 \right\}.
	\end{align*}
		Thus, $\beta \sim \text{N}(\mu, \Sigma)$ with parameters
		\begin{align*}
		\mu &= \Sigma \sum_{i=1}^n \left(y_i-\frac{1}{2}\right)x_i,\\
		\Sigma &= \left(2\sum_{i=1}^n \{\eta(\xi_i) x_i x_i^T\}+D\right)^{-1} \quad \text{with} \; D=\text{diag}((\mathbb{E}\gamma_{g(j)})_{j=1,\dots,p}).
		\end{align*}
		
		\paragraph{Relationship to the linear model}
				Note that the analogy to the linear update becomes explicit, when interpreting Equation (\ref{jaakkola_bound}) as a normal density on pseudo-data \citep{seeger} defined by
				$$\tilde{y}_{i} = \frac{2y_{i}-1}{4 \eta(\xi_{i})}.$$
				Then it can be easily seen that
				\begin{align*}
				\log h(\beta, \xi)= \log p(\tilde{y}|\beta) + c(\xi),
				\end{align*}
				where $$\tilde{y}_{i}|\beta \sim \text{N}(x_i^T\beta,(2\eta(\xi_i))^{-1}).$$ Replacing the precision parameter $\tau$ in the linear case with the precision of the pseudo-data (which is now sample-specific) can give us above updates directly from the linear model.
				
		\paragraph{Update for the variational parameter $\xi$}
		The update of the variational parameter $\xi$ is  given following \citep{Jaakkola} by
		$$\xi_i^2=x_i^T(\Sigma+ \mu_l\mu_l^T) x_i,$$
		which can be restricted to non-negative values of $\xi$ due to the symmetry.

		\paragraph{Evidence lower bound}
		As before 
			\begin{align*}
			\mathcal{L}(q)=\mathbb{E}_q \left(\log p(y, b, s, \gamma,  \pi)\right) + \sum_{i=1}^p H(q(b_i, s_i)) + H(q(\gamma))+ H(q(\pi)).
			\end{align*}
		The entropies can be calculated as in the linear model with the respective parameters of the variational distributions. The terms from the joint model density only differ in the first term
		\begin{align*}
		\mathbb{E}_q \log p(y, b , s , \gamma, \pi)&=\mathbb{E}_q \log p(y| b, s) + \mathbb{E}_q \log p(b| \gamma) \\
		& \qquad +\mathbb{E}_q \log p(s| \pi) +\mathbb{E}_q \log p(\gamma)+\mathbb{E}_q \log p(\pi),
		\end{align*}
		which here is given by 
		\begin{align*}
		\mathbb{E}_q \log p(y| \beta) &= \mathbb{E}_q\log \sigma((2y-1)\bfX\beta)\\
		&\geq \mathbb{E}_q \left(\frac{1}{2}\sum_i (2y_i-1) x_i^T \mu -\sum_i \eta(\xi_i) (x_i^T \beta)^2  \right. \\
		& \left. \qquad\qquad +\sum_i \left( \log(\sigma(\xi_i))-\frac{1}{2}\xi_i +\eta(\xi_i) \xi_i^2 \right)\right)	\\
	&=  \frac{1}{2} \sum_i \log(2\eta(\xi_i)) -\frac{1}{2}\sum_i 2\eta(\xi_i)  (\tilde{y}_i-x_i^T\mu)^2 +\text{const}.
		\end{align*}
	 This provides a lower bound on the evidence lower bound in analogy to the linear model that is used to monitor convergence.

		\subsection{Practical considerations}
		\label{appendix3}

			\subsubsection{Standardization of the predictors}
			\label{appendix_std}
\begin{figure}
			 	\centering
			 	\includegraphics[width=0.9\linewidth]{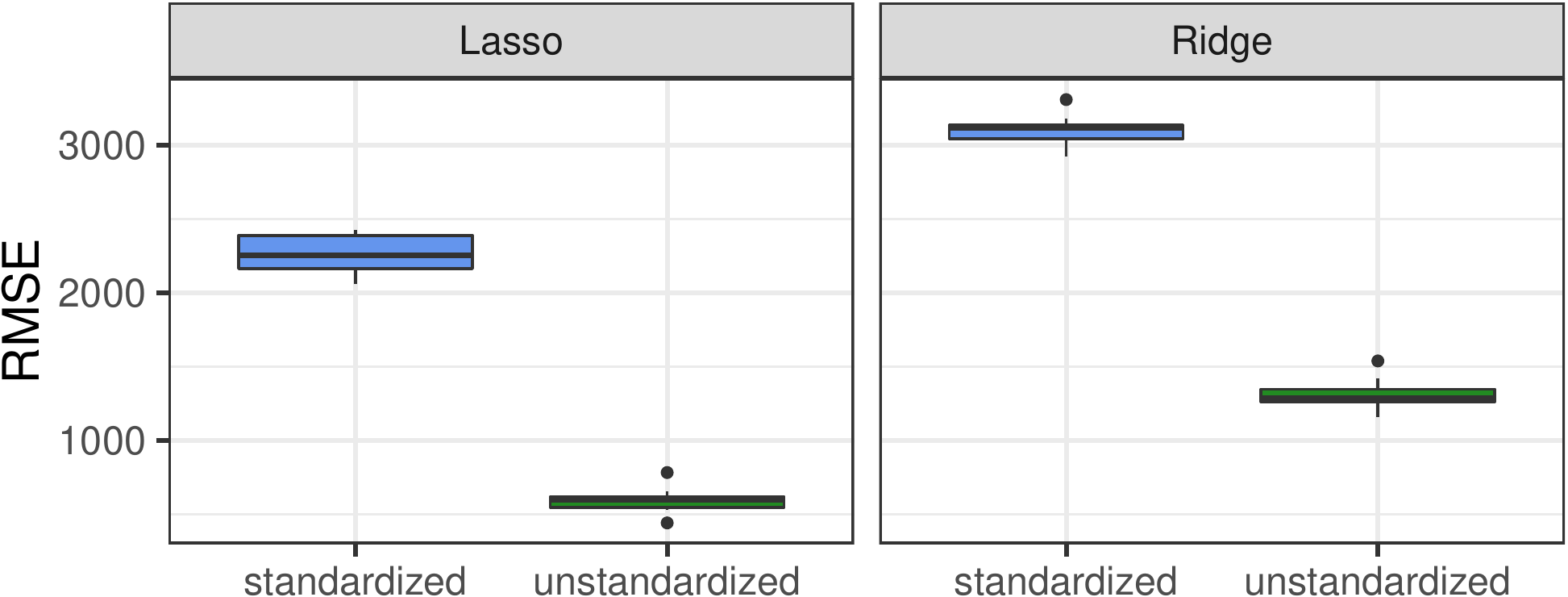}
			 	\caption{Simulation example illustrating the effect of standardization in settings with informative high-amplitude features and uninformative low-amplitude features. A number of $p=600$ features was simulated from a standard normal distribution and multiplied by 10 (high-amplitude features, $p=300$) or 1 (low-amplitude features, $p=300$). The response was simulated from a normal model with coefficients given by 1 for the high-amplitude features and 0 otherwise. Lasso and ridge regression were fitted on a training set of $n=500$ samples using either standardized predictors (blue) or predictors on the original scale (green). The resulting fits were evaluated in terms of the root mean squared error (RMSE) on an independent test set of $n=500$ samples. The boxplots were obtained from ten independent instances of simulated data.}
				\label{fig:std_example}
			 \end{figure}

		In penalised regression a common preprocessing step is the standardization of all predictors to unit variance to ensure a presumably 'fair' penalty. This scaling is in 1:1 correspondence to differential penalty factors. Without standardization features on a larger scale would be preferred as they need a smaller coefficient relative to a feature with the same effect but measured on a smaller scale. However, standardization can be suboptimal as it does not distinguish between meaningful differences in variance (e.g. features that differ between two disease groups) and differences in variance due to the scale. While removal of the latter would be desirable, meaningful differences should be retained. For example, in many applications we measure high-amplitude signals that are informative  jointly with low-amplitude features that originate mainly from technical noise. Here, standardization can be harmful (Figure \ref{fig:std_example}). Hence, the question of whether to scale the predictors or not, is related to the question of whether the variance of a feature is an informative covariate. 
		
		 By default, our method standardizes all features.
		 However, if we want to maintain the difference of variances within each assay, our method can adaptively learn scale differences between assays by $\gamma$, thereby removing the need to standardize for adjustment between assays  as seen in the CLL application.  This could help to retain meaningful differences in the features' variance within one assay. Alternatively, it is also possible to standardize features but re-include information on their variance via the covariate, e.g., binning features based on their variance. A recent study on RNA-Seq data found no strong effect of standardization compared to no standardization \citep{zwiener2014transforming}. Depending on the data set at hand it might, however, make sense to retain the original scale. For example with binary mutation data, where features are all on the same scale, standardization would favour mutations with lower frequencies.
			 
			\subsubsection{Modelling an intercept}
			To include an intercept into the model, we apply centering of $\bfX$ and $y$ before model fitting in the case of a linear model. For the logistic model this is not as straightforward and we follow \citep{carbonetto2012scalable} in the implementation, i.e. the intercept is $\beta_0$ is assumed to have a normal prior $\text{N}(0,\sigma_0^2)$ but considering the limiting case for $\sigma_0$ to infinity yielding an improper prior (essentially not penalizing the intercept).

\end{document}